\documentclass[a4paper,12pt,parskip,english]{scrartcl}
\setkomafont{disposition}{\normalfont\bfseries}
\usepackage{fullpage}
\usepackage{amsfonts}
\usepackage{hyperref}
\usepackage{makecell}
\usepackage{multirow}
\usepackage{placeins}
\usepackage{xcolor}
\usepackage{relsize}
\usepackage{cite} 
\usepackage{graphicx}
\usepackage[margin=1cm,font=footnotesize]{caption}
\usepackage[margin=0.5cm,font=scriptsize]{subcaption}
\usepackage{mathtools}
\usepackage{amssymb}
\usepackage{braket}

\newcommand{\figref}[1]{Fig. \ref{#1}}
\newcommand{\secref}[1]{Section \ref{#1}}
\newcommand{\secheadmath}[1]{\texorpdfstring{$#1$}{TEXT}} 
\numberwithin{equation}{section}

\newcommand{\pmatr}[1]{\begin{pmatrix} #1 \end{pmatrix}}
\newcommand{\phiatm}{\phi_{\mathrm{atm}}}
\newcommand{\phisol}{\phi_{\mathrm{sol}}}
\newcommand{\phidec}{\phi_{\mathrm{dec}}}
\newcommand{\simlt}{~\mbox{\smaller$\lesssim$}~}

\newcommand{\deltacp}{\delta_{\mathrm{CP}}}

\newcommand{\fit}{\chi^2}

\begin{document}
\title{\hfill ~\\[-5mm]
	\textbf{
		Testing constrained sequential dominance models of neutrinos}
}
\date{}
\author{\\[-5mm]
	Fredrik Bj\"{o}rkeroth\footnote{E-mail: {\tt F.Bjorkeroth@soton.ac.uk}}
	\ and Stephen F. King\footnote{E-mail: {\tt king@soton.ac.uk}}
	\\ \\
	\emph{\small School of Physics and Astronomy, University of Southampton,}\\
	\emph{\small Southampton, SO17 1BJ, United Kingdom}\\[4mm]
}

\maketitle

\begin{abstract}
{\noindent
Constrained sequential dominance (CSD) is a natural framework for implementing the see-saw mechanism of neutrino masses which allows the mixing angles and phases to be accurately predicted in terms of relatively few input parameters. We analyze a class of CSD($n$) models where, in the flavour basis, two right-handed neutrinos are dominantly responsible for the ``atmospheric'' and ``solar'' neutrino masses with Yukawa couplings to $(\nu_e, \nu_{\mu}, \nu_{\tau})$ proportional to $(0,1,1)$ and $(1,n,n-2)$, respectively, where $n$ is a positive integer. These coupling patterns may arise in indirect family symmetry models based on $A_4$. With two right-handed neutrinos, using a $\chi^2$ test, we find a good agreement with data for CSD(3) and CSD(4) where the entire PMNS mixing matrix is controlled by a single phase $\eta$, which takes simple values, leading to accurate predictions for mixing angles and the magnitude of the oscillation phase $|\deltacp|$. We carefully study the perturbing effect of a third ``decoupled'' right-handed neutrino, leading to a bound on the lightest physical neutrino mass $m_1\simlt 1$ meV for the viable cases, corresponding to a normal neutrino mass hierarchy. We also discuss a direct link between the oscillation phase $\deltacp $ and leptogenesis in CSD($n$) due to the same see-saw phase $\eta$ appearing in both the neutrino mass matrix and leptogenesis.
} 
\end{abstract}
\thispagestyle{empty}
\vfill
\newpage
\setcounter{page}{1}

\section{Introduction}
The astonishingly accurate measurement of the third lepton mixing angle, the so-called reactor angle $\theta_{13}\approx 8.5^{\circ}\pm 0.2^{\circ}$ \cite{An:2012eh}, signals the start of the precision era for neutrino physics. Over the coming years, all three lepton mixing angles are expected to be measured with increasing precision. A first tentative hint for a value of the CP-violating phase $\deltacp \sim -\pi /2$ has also been reported in global fits \cite{Gonzalez-Garcia:2014bfa,Forero:2014bxa,Capozzi:2013csa}. However the mass squared ordering (normal or inverted), the scale (mass of the lightest neutrino) and nature (Dirac or Majorana) of neutrino mass so far all remain unknown.%
\footnote{The first two attributes are commonly referred to jointly as the ``mass hierarchy'', although really they are separate questions.}%

On the theory side, there are many possibilities for the origin of light neutrino masses $m_i$ and mixing angles $\theta_{ij}$. Perhaps the simplest and most elegant idea is the classical see-saw mechanism, in which the observed smallness of neutrino masses is due to the heaviness of right-handed Majorana neutrinos \cite{Minkowski:1977sc},
\begin{equation}
m^{\nu}=m^DM^{-1}_{R}(m^D)^T,
\label{seesaw}
\end{equation}
where $m^{\nu}$ is the light effective left-handed%
\footnote{We have ignored the overall physically irrelevant phase of -1.}
Majorana neutrino mass matrix (i.e. the physical neutrino mass matrix), $m^D$ is the Dirac mass matrix (in LR convention) and $M_R$ is the (heavy) Majorana mass matrix. Although the see-saw mechanism generally predicts Majorana neutrinos, it does not predict the ``mass hierarchy'', nor does it yield any understanding of lepton mixing. In order to overcome these deficiencies, the see-saw mechanism must be supplemented by other ingredients. 

\begin{figure}[t]
\centering
\includegraphics[width=0.4\textwidth]{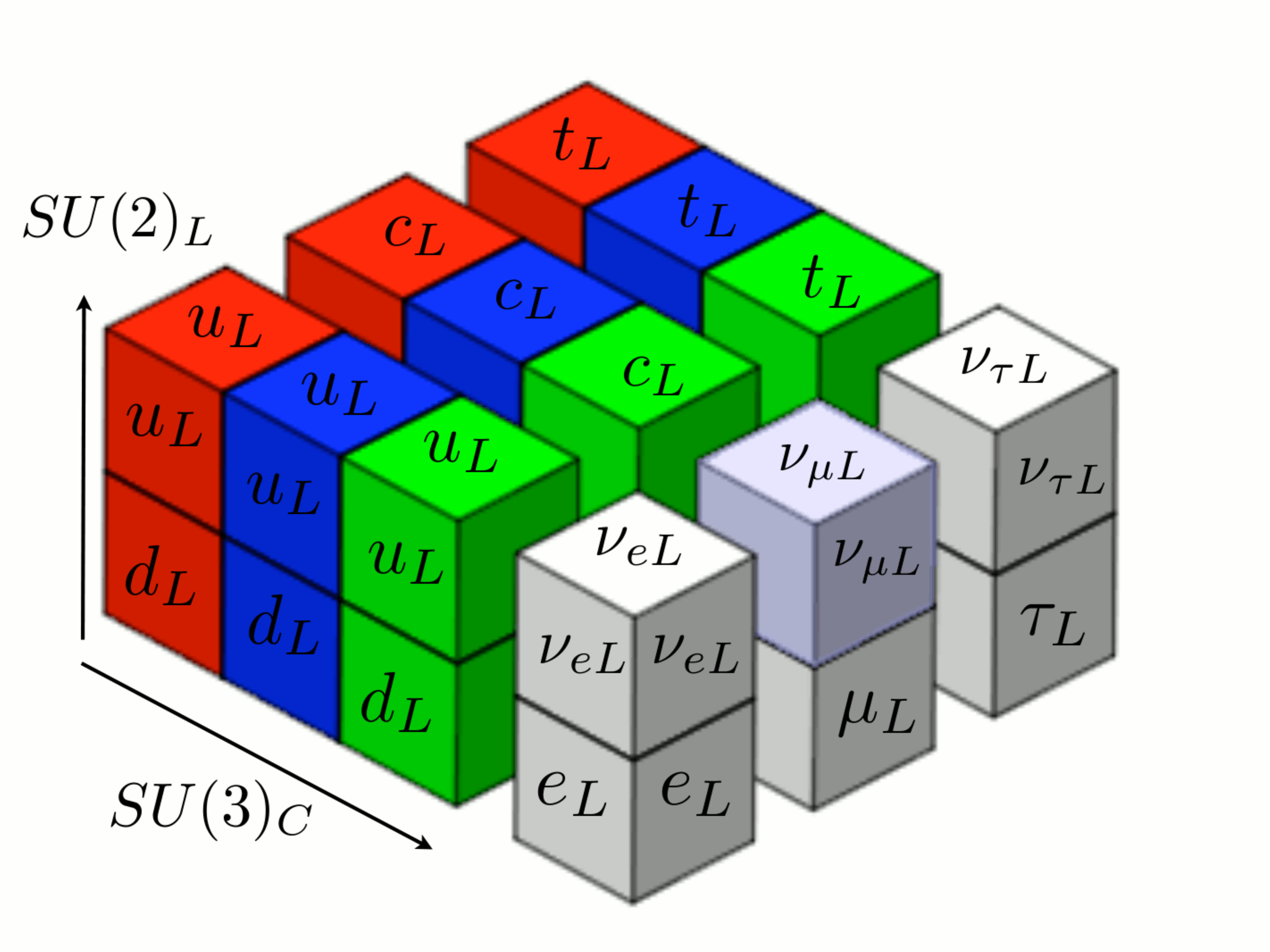}
\includegraphics[width=0.4\textwidth]{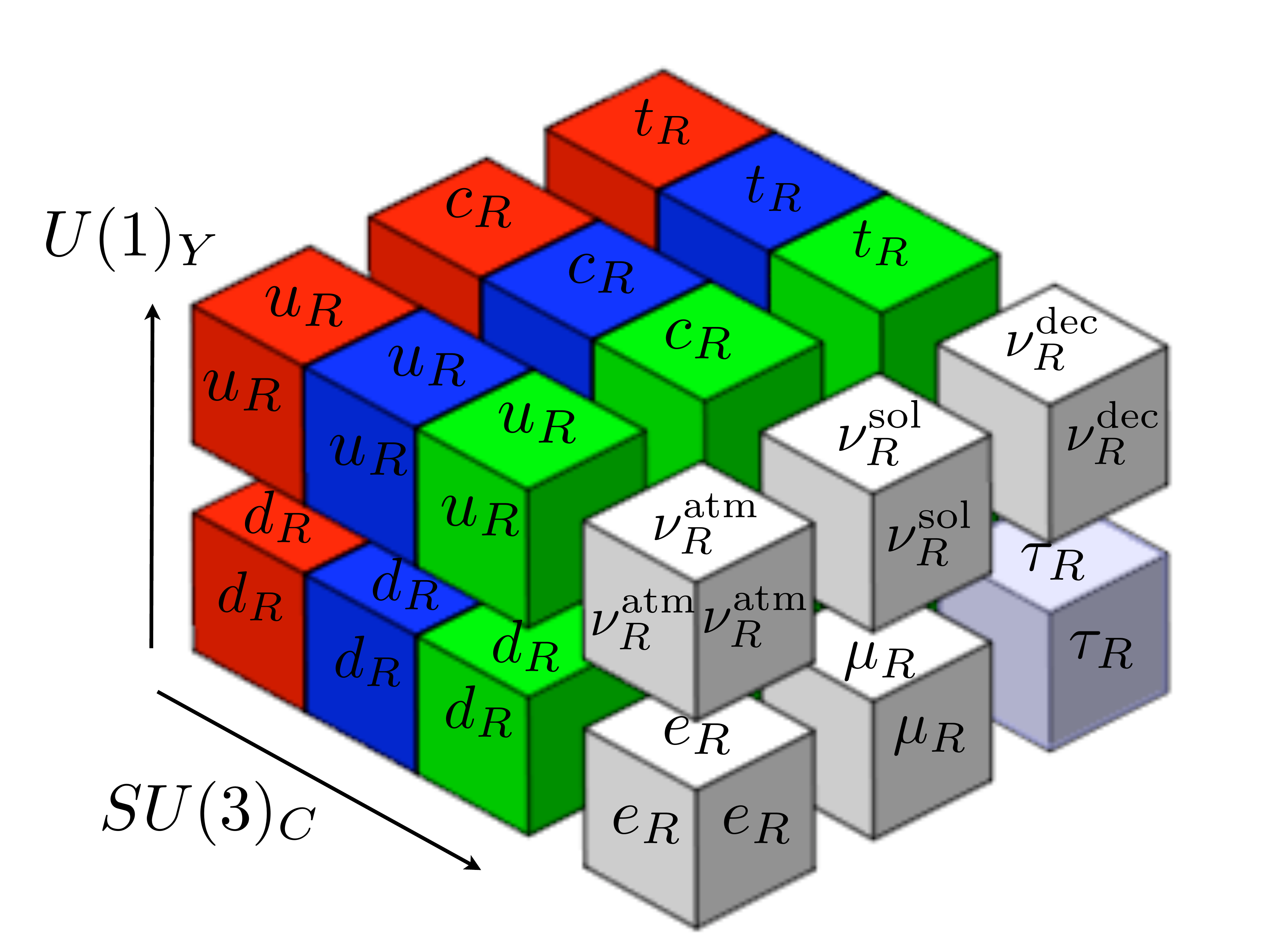}
\vspace*{-4mm}
\caption{The Standard Model (SM) with three right-handed neutrinos defined as $(\nu_R^{\rm atm},\nu_R^{\rm sol},\nu_R^{\rm dec})$ which in sequential dominance are mainly responsible for the $m_3,m_2,m_1$ physical neutrino masses, respectively.} 
\label{SM}
\vspace*{-2mm}
\end{figure}

One attractive idea, depicted in \figref{SM}, is that the Standard Model (SM) is supplemented by three right-handed neutrinos which contribute sequentially to the light effective neutrino mass matrix. The idea of such a ``sequential dominance'' (SD) \cite{King:1998jw} is that one dominant right-handed neutrino $\nu^{\rm atm}_R$ of mass $M_{\rm atm}$ is mainly responsible for the heaviest atmospheric neutrino mass $m_3$, while a second subdominant right-handed neutrino $\nu^{\rm sol}_R$ of mass $M_{\rm sol}$ mainly gives the solar neutrino mass $m_2$. A third, approximately decoupled, right-handed neutrino $\nu^{\rm dec}_R$ of mass $M_{\rm dec}$ is responsible for the lightest neutrino mass $m_1$. In the diagonal basis, $M_R={\rm diag}(M_{\rm atm},M_{\rm sol},M_{\rm dec})$ where the Dirac mass matrix is constructed from three columns $m^D=(m^D_{\rm atm},m^D_{\rm sol},m^D_{\rm dec})$, applying the see-saw formula in Eq.~\ref{seesaw} gives,
\begin{equation}
m^{\nu}=\frac{m^D_{\rm atm}(m^D_{\rm atm})^T}{M_{\rm atm}}
+
\frac{m^D_{\rm sol}(m^D_{\rm sol})^T}{M_{\rm sol}}
+
\frac{m^D_{\rm dec}(m^D_{\rm dec})^T}{M_{\rm dec}},
\end{equation}
where 
\begin{equation}
\frac{(m^D_{\rm atm})^{\dagger}m^D_{\rm atm}}{M_{\rm atm}}
>\frac{(m^D_{\rm sol})^{\dagger}m^D_{\rm sol}}{M_{\rm sol}}
\gg
\frac{(m^D_{\rm dec})^{\dagger}m^D_{\rm dec}}{M_{\rm dec}},
\end{equation}
leading immediately to the prediction of a normal mass hierarchy of physical neutrino masses $m_3> m_2\gg m_1$.

The observed pattern of lepton mixing angles can be understood in the above SD framework as follows. In the diagonal charged lepton and right-handed neutrino mass basis, if the dominant ``atmospheric'' right-handed neutrino  $\nu_R^{\rm atm}$ has couplings $(m^D_{\rm atm})^T=(0,a_1,a_2)$ to $(\nu_e, \nu_{\mu}, \nu_{\tau})$, then this implies $\tan \theta_{23}\sim a_1/a_2$ \cite{King:1998jw} and a bound $\theta_{13} \simlt m_2/m_3$ \cite{King:2002nf}. The subdominant ``solar'' right-handed neutrino  $\nu_R^{\rm sol}$ couplings $(m^D_{\rm sol})^T=(b_1,b_2,b_3)$ to $(\nu_e, \nu_{\mu}, \nu_{\tau})$ further yield $\tan \theta_{12} \sim \sqrt{2}b_1/(b_2-b_3)$ \cite{King:1998jw,King:2002nf}. However in practice these estimates are subject to large corrections beyond the SD approximation, and as we shall see later, the atmospheric and reactor angle in particular depend sensitively on a choice of phase. The lepton mixing angles are of course insensitive to the ``decoupled'' right-handed neutrino couplings.

In order to obtain sharp predictions for lepton mixing angles, the relevant Yukawa coupling ratios need to be fixed, for example using vacuum alignment of family symmetry breaking flavons (for reviews see e.g.~\cite{Altarelli:2010gt,Ishimori:2010au,King:2013eh,King:2014nza}). The first attempt to use vacuum alignment within an $SU(3)$ family symmetry to predict maximal atmospheric mixing ($\tan \theta_{23} \sim 1$) from equal dominant right-handed neutrino couplings $(m^D_{\rm atm})^T=(0,a,a)$ was discussed in \cite{King:2001uz}. Subsequently, constrained sequential dominance (CSD) \cite{King:2005bj} was proposed to explain tri-bimaximal (TB) mixing with a zero reactor angle by using vacuum alignment to fix the subdominant ``solar'' right-handed neutrino couplings to $(\nu_e, \nu_{\mu}, \nu_{\tau})$ to also be equal up to a sign,%
\footnote{Note that $(0,a,a)\cdot(b,b,-b)=0$. This orthogonality is related to the fact that CSD(1) respects form dominance, since columns of the Dirac mass matrix in the flavour basis are proportional to the columns of the unitary PMNS mixing matrix \cite{Chen:2009um}.}
namely $(m^D_{\rm sol})^T=(b,b,-b)$.

From the point of view of discrete family symmetry models, the above approach is sometimes referred to as ``indirect'' since the required vacuum alignments completely break the family symmetry \cite{King:2009ap}. Such ``indirect'' models are highly predictive and do not require such large discrete groups as the ``direct'' models which use only vacuum alignments which preserve a subgroup $\mathbb{Z}_2 \times \mathbb{Z}_2$ in the neutrino sector (the so-called Klein symmetry) and  $\mathbb{Z}_3$ in the charged lepton sector, where such an approach requires $\Delta (6N^2)$ for large values of $N$ \cite{Holthausen:2012wt,King:2013vna,Lavoura:2014kwa,Fonseca:2014koa}.%
\footnote{An analogous approach based on $\Delta (6N^2)$ has also been considered in the quark sector \cite{Araki:2013rkf}.}

In this paper we perform a dedicated analysis of the general class of CSD($n$) models, independently of any detailed model, allowing the positive integer $n$ to take any value. Thus we consider the 
dominant ``atmospheric'' right-handed neutrino and the subdominant ``solar'' right-handed neutrino to have couplings to $(\nu_e, \nu_{\mu}, \nu_{\tau})$ given by:
\begin{equation}
\setlength{\itemsep}{-1.5ex}
{\rm CSD}(n): \qquad (m^D_{\rm atm})^T=(0,a,a), \ \ \ \ (m^D_{\rm sol})^T=(b,nb,(n-2)b),\hspace{10ex}
\label{sol}
\end{equation}
where $n$ is any positive integer. This is a generalisation of examples studied in the literature so far for $n=2,3,4$, including the original CSD identified here as CSD(1). After the see-saw mechanism has been implemented, with just two right-handed neutrinos, the light effective Majorana neutrino mass matrix depends on just two mass parameters $ m_a $ and $ m_b $ and a relative phase $ \eta $ 
(three real input parameters). 
For each value of $ n $ we perform a fit to five observed neutrino parameters: three mixing angles and two mass squared differences.


We find good fits for the CSD(3) and CSD(4) parameters, with favoured values of $ \eta $ near $2\pi/3$ and $4\pi/5$, respectively, consistent with spontaneous CP violation of an Abelian symmetry $\mathbb{Z}_{3N}$ or $\mathbb{Z}_{5N}$ symmetry, as previously observed \cite{King:2013iva,King:2013xba}. Unlike these earlier studies, however, here we perform a global fit leading to more robust results which allow the input phase to be determined from the data on the mixing angles. Indeed it is reassuring to see the simple rational values of the input phase $2\pi/3$ or $4\pi/5$ emerge from the fit.

The value of the CP phase $\deltacp$ emerges as a genuine prediction. Moreover, with just two right-handed neutrinos in CSD($n$), there is a direct link between the oscillation phase $\deltacp $ and the leptogenesis phase since there is only one phase $\eta$ in the see-saw matrices which is responsible for both. The more general case with a third approximately decoupled right-handed neutrino provides a close approximation to this situation. Therefore in both cases, observation of leptonic CP violation in low energy neutrino oscillation experiments is directly linked to cosmological CP violation, which both vanish in the same limit.

We shall consider the effect of a third almost decoupled right-handed neutrino giving $ m^D_{\rm dec} \propto (0,0,1) $, which introduces a further mass parameter $m_c$ and relative phase $\xi$, in order to gauge the effect of having a non-zero lightest neutrino mass $m_1$. For low values of $m_c$, this provides a perturbation to our previous results leading to an upper limit on the lightest physical neutrino mass $m_1\simlt 1$ meV for the viable cases.

Though our analysis here is independent of a specific model (such as a GUT), it is to be understood that the CSD alignments are discussed with a mind to integration within a more complete model that ideally can explain all fermionic mass and mixing. As such, numerical results presented here give an important foundational step in an approach to solving the flavour puzzle.

The remainder of the paper is set out as follows. 
\secref{sec:conventions} defines the see-saw conventions and gives the CSD($n$) neutrino mass matrices.
In \secref{sec:A4} we discuss a generic flavour model framework, based on $A_4$ family symmetry, where the CSD(n) vacuum alignments arise due to orthogonality conditions between flavons.
In \secref{sec:chisq} we define a test-statistic $ \fit $ and outline the method for finding and evaluating the global minima in parameter space. In \secref{sec:results} we present the results of our $ \fit $ analysis, first for the case of two right-handed neutrinos, then with three right-handed neutrinos, plotting the results against $ n $, the input phase $ \eta $, and the mass of the lightest neutrino $ m_1 $, for two choices of relative phase between submatrices of the neutrino mass matrix. 
In subsection \ref{sec:special} we consider particular phase choices for CSD(3) and CSD(4), prompted by the results of our fit and specific model considerations.
In \secref{link} we discuss the link between the oscillation phase $\deltacp $ and leptogenesis in CSD($n$).
\secref{conclusion} concludes the paper.
Appendix \ref{sec:chisqinput} discusses in more detail the  $ \fit $ distribution of the parameters.

\section{See-saw conventions and CSD(\secheadmath{n}) mass matrices}
\label{sec:conventions}
\begingroup
The charged lepton and neutrino Yukawa matrices $Y^{e}$, $Y^{\nu}$ are defined in a LR convention by%
\footnote{This LR convention for the Yukawa matrix differs by an Hermitian conjugation compared to that used in the MPT package \cite{Antusch:2005gp} due to the RL convention used there.}
\begin{equation}
	\mathcal{L}^{LR} = 
	-H^dY^e_{ij}\overline L^i_{\mathrm{L}} e^j_{\mathrm{R}} - H^u Y^{\nu}_{ij}
	\overline L^i_{\mathrm{L}}  \nu^{i}_{R} + \mathrm{h.c.}
\label{Ynu}
\end{equation}
where $i,j=1,2,3$ label the three families of lepton doublets $L_i$, right-handed charged leptons $e^j_R$ and right-handed neutrinos $\nu_R^j$; $ H^u $, $H^d$ are the electroweak Higgs doublets which develop VEVs $v_u,v_d$. The physical effective neutrino Majorana mass matrix $m^{\nu}$ is determined from the columns of $Y^{\nu}$ via the see-saw mechanism,
\begin{equation}
	m^{\nu} = v_u^2 Y^{\nu} M^{-1}_{RR} Y^{\nu \mathrm{T}},
\label{eq:seesaw}
\end{equation}
where the light Majorana neutrino mass matrix $m^\nu$ is defined%
\footnote{Note that this convention for the light effective Majorana neutrino mass matrix $m^{\nu}$ differs by an overall complex conjugation compared to that used in the MPT package \cite{Antusch:2005gp}.}
by $ \mathcal{L}^{LL}_\nu = -\tfrac{1}{2} m^\nu \overline{\nu}_{L} \nu^{c}_L + \mathrm{h.c.} $, while the heavy right-handed Majorana neutrino mass matrix $M_R$ is defined by $\mathcal{L}^{RR}_\nu = -\tfrac{1}{2} M_{RR} \overline{\nu^c}_{R} \nu_{R} + \mathrm{h.c.} $.

In the above conventions, the CSD($n$) mass matrices are defined as in Eq.~\ref{sol},
\begin{equation}
m^D = 
	Y^{\nu}v_u=\pmatr{0 & b & 0 \\ a & nb & 0 \\ a & (n-2)b & c}, \quad
	M_{R} = \pmatr{M_{\rm atm} & 0 & 0 \\ 0 & M_{\rm sol}  & 0 \\ 0 & 0 & M_{\rm dec}}.
	\label{CSDn}
\end{equation}
Applying the see-saw in Eq.~\ref{seesaw} for these matrices gives a light neutrino mass matrix,
\begin{equation}
	m^\nu_{(n)} = m_a e^{i\alpha} \pmatr{0&0&0\\0&1&1\\0&1&1} + m_b e^{i\beta} \pmatr{1&n&n-2\\n&n^2&n(n-2)\\n-2&n(n-2)&(n-2)^2} + m_c e^{i\gamma} \pmatr{0&0&0\\0&0&0\\0&0&1},
	\label{eq:mnu}
\end{equation}
where $ m_a=|a|^2 /M_{\rm atm} $, $ m_b = |b|^2 /M_{\rm sol}$ and $ m_c = |c|^2 /M_{\rm dec}$ are real and positive combinations of other physical parameters, with the phases displayed explicitly as $ \alpha= \arg (a^2 ) $, $ \beta = \arg (b^2 )$ and $ \gamma = \arg (c^2)$. An overall unphysical phase $\alpha$ may be factored out and then dropped in order to make the term proportional to $m_a$ is real, wherein we make the redefinitions $ \eta = \beta - \alpha $ and $ \xi = \gamma - \alpha$. Hence $\eta = \arg(b^2/a^2)$ and $\xi = \arg(c^2/a^2)$.

The neutrino mass matrix $m^\nu$ is diagonalised by
\begin{equation}
	U_{\nu_L} m^\nu U^T_{\nu_L} = \pmatr{m_1 & 0 & \\ 0 & m_2 & 0 \\ 0 & 0 & m_3}.
\end{equation}
The PMNS matrix is then given by $U_{\mathrm{PMNS}} = U_{e_L} U^\dagger_{\nu_L} $, where $U_{e_L}$ is given by
\begin{equation}
	U_{e_L} Y^e U^\dagger_{e_R} = \pmatr{y_e & 0 & 0 \\ 0 & y_\mu & 0 \\ 0 & 0 & y_\tau}.
\end{equation}
We use the standard PDG parameterization \cite{pdg} $U_{\mathrm{PMNS}} = R^l_{23} U^l_{13} R^l_{12} P_\textrm{PDG}  $ in terms of $s_{ij}=\sin \theta_{ij}$, $c_{ij}=\cos\theta_{ij}$, the Dirac CP violating phase $\deltacp$ and further Majorana phases contained in  $P_\textrm{PDG} = \textrm{diag}(1,e^{i\frac{\alpha_{21}}{2}},e^{i\frac{\alpha_{31}}{2}})$. We shall assume that $Y^e$ is diagonal, hence $ U_{e_L} $ is the identity matrix up to diagonal phase rotations, and that $ U_\mathrm{PMNS} = U^\dagger_{\nu_L} $, i.e. simply the matrix that diagonalises the neutrino mass matrix, up to charged lepton phase rotations. 
We will now show how a diagonal charged lepton Yukawa matrix and a neutrino Yukawa matrix with CSD($n$) structure can be achieved in a generic model based on $ A_4 $ family symmetry.

\section{CSD(\secheadmath{n}) from \secheadmath{A_4}}
\label{sec:A4}

Following the measurement of the reactor angle, various types of CSD have been proposed, with the dominant right-handed ``atmospheric'' couplings as above,
\begin{equation}
(m^D_{\rm atm})^T=(0,a,a),
\label{atm2}
\end{equation}
and hence an approximate maximal atmospheric angle $\tan \theta_{23}\sim a_1/a_2 \sim 1$, while proposing alternative subdominant ``solar'' right-handed neutrino couplings as follows:
\begin{itemize}
\setlength{\itemsep}{-1.5ex}
\item CSD(2): $(m^D_{\rm sol})^T=(b,2b,0)$ \cite{Antusch:2011ic}.
\item CSD(3): $(m^D_{\rm sol})^T=(b,3b,b)$ \cite{King:2013iva}.
\item CSD(4): $(m^D_{\rm sol})^T=(b,4b,2b)$ \cite{King:2013iva,King:2013xba}.
\end{itemize}
All these examples maintain an approximate trimaximal value for the solar leptonic angle $\tan \theta_{12} \sim \sqrt{2}b_1/(b_2-b_3) \sim 1/\sqrt{2}$, while switching on the reactor angle. Since experiment indicates that the bound $\theta_{13} \simlt m_2/m_3$ is almost saturated, these schemes also require certain phase choices $\arg (b/a)$ in order to achieve the desired reactor angle, leading to predictions for the CP-violating phase $\deltacp$.%
\footnote{Note that CSD(4), when implemented in unified models with $Y^u=Y^{\nu}$, with the second column proportional to $(1,4,2)$, predicts a Cabibbo angle $\theta_C\approx 1/4$ in the diagonal $Y^d\sim Y^e$ basis. Pati-Salam models have been constructed along these lines \cite{King:2013hoa}.}
Our goal in this paper is generalise and then systematically study such patterns of couplings,
which we refer to as CSD($n$) defined in Eq.\ref{sol}. But first we should justify such a pattern of couplings
and show how it may arise from a more fundamental theory based on a non-Abelian family symmetry.

The basic starting point is to consider some family symmetry such as $A_4$ which admits triplet representations. The family symmetry is broken by triplet flavons $\phi_i$ whose vacuum alignment will control the structure of the Yukawa couplings. Consider for example a supersymmetric model, where the relevant superpotential terms that produce the correct Yukawa structure in the neutrino sector are
\begin{equation}
	\frac{1}{\Lambda} H_u (L \cdot \phiatm) \nu_{\rm atm}^c + \frac{1}{\Lambda} H_u (L \cdot\phisol )
	\nu_{\rm sol}^c + \frac{1}{\Lambda} H_u (L \cdot\phidec)\nu_{\rm dec}^c ,
\label{Ynu_flavon}
\end{equation}
where $ L $ is the SU(2) lepton doublet, assumed to transform as a triplet under the family symmetry, while 
$ \nu_{\rm atm}^c, \nu_{\rm sol}^c, \nu_{\rm dec}^c$ are CP conjugates of the 
right-handed neutrinos and $ H_u $ is the electroweak scale up-type Higgs field, the latter being family symmetry singlets but distinguished by some additional quantum numbers. In the charged-lepton sector,
\begin{equation}
	\frac{1}{\Lambda} H_d (L \cdot \phi_e) e^c  + \frac{1}{\Lambda} H_d (L \cdot \phi_\mu) \mu^c + \frac{1}{\Lambda} H_d (L \cdot\phi_\tau) \tau^c ,
\end{equation}
where $ e^c,\mu^c,\tau^c$ are the CP conjugated right-handed electron, muon and tau respectively. The right-handed neutrino Majorana superpotential is typically chosen to give a diagonal mass matrix,
\begin{equation}
	M_{R} = \mathrm{diag}(M_{\rm atm},M_{\rm sol},M_{\rm dec}).
	\label{eq:mrr}
\end{equation}
Details of the construction of this superpotential (e.g. in terms of flavons), the relative values of 
$M_{\rm atm},M_{\rm sol},M_{\rm dec}$
as well as the inclusion of any off-diagonal terms in $ M_{R} $ will all depend on the additional specifications of the model. 

The CSD($n$) vacuum alignments arise from effective operators involving three flavon fields $ \phiatm $, $ \phisol $, and $ \phidec $ which are triplets under the flavour symmetry and acquire VEVs. The subscripts are chosen by noting that $ \phiatm $ correlates with the atmospheric neutrino mass $ m_3 $, $ \phisol $ with the solar neutrino mass $ m_2 $, and $ \phidec $ with the lightest neutrino mass $ m_1 $, which in CSD is light enough that the associated third right-handed neutrino can, to good approximation, be thought of as decoupled from the theory \cite{King:1998jw}. CSD($n$) is defined to be the choice of vacuum alignments,
\begin{equation}
	\braket{\phiatm} \propto \pmatr{0 \\ 1 \\ 1}, \qquad \braket{\phisol} \propto \pmatr{1\\n\\n-2}, \qquad \braket{\phidec} \propto \pmatr{0\\0\\1},
	\label{CSD(n)}
\end{equation}
where $n$ is a positive integer, and the only phases allowed are in the overall proportionality constants.%
\footnote{In general also the elements of flavon VEVs can have relative signs
as in the last alignment in Eq.~\ref{sym}. However, for a given choice of such alignment, orthogonality fixes the relative signs of the elements of subsequent alignments with only an overall complex proportionality factor remaining.}
Such vacuum alignments arise from symmetry preserving alignments together with orthogonality conditions  
\cite{King:2013iva,King:2013xba}, as discussed below.

The starting point for understanding the alignments in Eq.~\ref{CSD(n)} are the symmetry preserving vacuum alignments of $A_4$,
namely:
\begin{equation}
	\pmatr{1 \\ 0 \\ 0} ,  \pmatr{0\\1\\0} , \pmatr{0\\0\\1} , \pmatr{\pm 1\\ \pm 1\\ \pm 1},
	\label{sym}
\end{equation}
which each preserve some subgroup of $A_4$ in a basis where the 12 group elements in the triplet representation are real
(i.e. each alignment in Eq.~\ref{sym} is an eigenvector of at least one non-trivial group element with eigenvalue +1).
In a flavour model the above alignments would also arise from the VEVs of triplet flavons, which however do not couple to fermions. As such, their immediate role beyond producing the CSD($ n $) alignments is unclear, though they may have an impact on early universe physics, for example
in flavon inflation \cite{Antusch:2008gw}.
The first alignment in Eq.~\ref{CSD(n)}, which completely breaks the $A_4$ symmetry,
arises from the orthogonality conditions 
\begin{equation}
\begin{pmatrix}0 \\ 1\\ 1\end{pmatrix}
\perp 
\begin{pmatrix}1 \\1\\-1\end{pmatrix},
\begin{pmatrix}1 \\0\\0\end{pmatrix}
\label{011}
\end{equation}
involving two symmetry preserving alignments selected from Eq.~\ref{sym}.
The following symmetry breaking alignment may be obtained which is orthogonal to the alignment in Eq.~\ref{011} and one of the 
symmetry preserving alignments, 
\begin{equation}
\begin{pmatrix}2 \\-1\\1\end{pmatrix}
\perp 
\begin{pmatrix}1 \\1\\-1\end{pmatrix},
\begin{pmatrix}0 \\1\\1\end{pmatrix}
\label{2,-1,1}
\end{equation}
The CSD($n$) alignment in Eq.~\ref{CSD(n)} is orthogonal to the above alignment in Eq.~\ref{2,-1,1},
\begin{equation}
\begin{pmatrix}1 \\n\\ n-2 \end{pmatrix}
\perp 
\begin{pmatrix}2 \\-1\\1\end{pmatrix}
\label{CSD(n)again}
\end{equation}
where the orthogonality in Eq.~\ref{CSD(n)again} is maintained for any value of $n$ (not necessarily integer).
To pin down the value of $n$ and show that it is a particular integer requires a further orthogonality condition.%
\footnote{We could simply use the alignment in Eq.~\ref{CSD(n)again}, where $ n $ is a real number to be fitted. However, we prefer to fix $n$ to be a small positive integer
to increase predictivity.}

For example, for $n=3$, the desired alignment is obtained from the two orthogonality conditions,
\begin{equation}
\begin{pmatrix}1 \\3\\1\end{pmatrix}
\perp 
\begin{pmatrix}2 \\-1\\1\end{pmatrix},
\begin{pmatrix}1 \\0\\-1\end{pmatrix}
\label{CSD(3)again}
\end{equation}
where the first condition above is a particular case of Eq.~\ref{CSD(n)again}
and the second condition involves a new alignment,
obtained from two of the symmetry preserving alignments in Eq.~\ref{sym}, 
\begin{equation}
\begin{pmatrix}1 \\0\\-1\end{pmatrix}
\perp 
\begin{pmatrix}1 \\1\\1\end{pmatrix},
\begin{pmatrix}0 \\1\\0\end{pmatrix}
\label{1,0,-1}
\end{equation}

Using Eq.~\ref{Ynu_flavon}, the vacuum alignments in Eq.~\ref{CSD(n)} make up the columns of the Dirac neutrino Yukawa matrix 
$ Y^{\nu} \propto (\braket{\phiatm}, \braket{\phisol}, \braket{\phidec}) $, giving a Dirac mass matrix
\begin{equation}
	m^D = 
	Y^{\nu}v_u=\pmatr{0 & b & 0 \\ a & nb & 0 \\ a & (n-2)b & c}, 	\label{eq:ynu}
\end{equation}
which is consistent with Eq.~\ref{sol}
where $m^D=(m^D_{\rm atm},m^D_{\rm sol},m^D_{\rm dec})$
and the coefficients $ a $, $ b $, and $ c $ are generally complex. The charged-lepton Yukawa matrix is chosen to be diagonal (up to model-dependent corrections, assumed small), corresponding to the existence of three flavons $ \phi_e $, $ \phi_\mu $ and $ \phi_\tau $ in the charged-lepton sector which acquire VEVs with alignments \cite{King:2013iva,King:2013xba}
\begin{align}
	\braket{\phi_e} & \propto \pmatr{1 \\ 0 \\ 0} & \braket{\phi_\mu} & \propto \pmatr{0\\1\\0} & \braket{\phi_\tau} & \propto \pmatr{0\\0\\1}.
\end{align}
Given this choice, it is clear that $Y^e$ is diagonal, hence $ U_{e_L} $ is the identity matrix up to diagonal phase rotations, and that $ U_\mathrm{PMNS} = U^\dagger_{\nu_L} $, i.e. simply the matrix that diagonalises the neutrino mass matrix, up to charged lepton phase rotations.

\section{Our fitting method}
\label{sec:chisq}
\subsection{The test function \secheadmath{\fit}}
We first clarify that we do not use raw experimental data. Instead our ``data'' corresponds to global fit values $\mu_i$ to true experimental data, where 
$$
\mu_i\in \{\sin^2\theta_{12}, \sin^2\theta_{23}, \sin^2\theta_{13}, \Delta m_{21}^2,  \Delta m_{31}^2\}.
$$
We are not performing a global fit to the data, but instead are fitting our model parameters, which are collected into a vector $x$, to the existing results of a global fit. For each value of the input vector $x$ we obtain a set of predicted values $P_i$ which may be compared to the global best fit values $\mu_i$. We want to use the test function $\chi^2$ to determine the optimum input parameters $x$ corresponding to the predictions $P_i$ which yield the lowest $\chi^2$. For definiteness, all ``data'' $\mu_i$ is taken from just one of the global fits, namely that in \cite{Gonzalez-Garcia:2014bfa}, which is reproduced in Table \ref{tab:data} for the case of a normal mass squared ordering predicted by CSD($n$) models. Bounds exist for the CP-violating phase $\delta_{\mathrm{CP}}$ at 1$\sigma$, but it is completely undetermined at 3$\sigma$, and so is left as a pure prediction, as are the two Majorana phases.

We define the function $\fit$ to serve as a test-statistic for the goodness-of-fit of a chosen vector $x=(m_a,m_b,m_c,\eta,\xi)$ in input parameter space in analogy to \cite{Feruglio:2014jla},
\begin{equation}
	\fit = \sum_{i=1}^N \left( \frac{P_i(x)-\mu_i}{\sigma_i}\right)^2,
	\label{eq:chisq}
\end{equation}
where $ N = 5 $, $\mu_i$ are the current global best fit values to the experimental data, and $\sigma_i$ are the 1$\sigma$ deviations for each of the five physical predictions $P_i$ made, i.e. for the (squared sines of) three PMNS mixing angles $\theta_{ij}$ and two mass-squared differences $\Delta m_{21}^2$ and $\Delta m_{31}^2$. Note that $ \sigma_i $ is equivalent to the standard deviation of the global best fit values if the global fit distribution of the observable is Gaussian. This is essentially the case for most fitted observables, except for the atmospheric angle $ \theta_{23} $. As seen in \cite{Gonzalez-Garcia:2014bfa}, the $ \Delta\chi^2 $ distribution for $ \theta_{23} $ has two minima on either side of $ 45^\circ $, with a slight preference for $ \theta_{23} < 45^\circ $ for Normal Ordering ($ 42.3^\circ $, to be precise). This is reflected in the asymmetric error $ ^{+3.0^{\circ}}_{-1.6^{\circ}} $ which in terms of $\sin^2 \theta_{23}$ is $ ^{+0.052}_{-0.028} $ (see Table \ref{tab:data} below). So as to not overstate the error (and consequently underestimate $ \fit $), we assume the distribution is Gaussian about the best fit, setting $ \sigma_{\sin^2\theta_{23}} = 0.028$. For best fit values larger than $ 42.3^\circ $ this will overestimate the $\chi^2 $, so we are being very conservative in presenting our results when the global fit ``data'' we are using is not Gaussian.

One may compare the resulting values of $\chi^2$ to the number of excess degrees of freedom $ \nu \equiv N-N_I $ where $ N=5 $ is the number of ``data points'' (in our case interpreted as the number of measured parameters) while $ N_I $ is the number of input parameters. In standard $ \fit $ analyses, the $ \fit $ per excess degree of freedom for a good fit has an expectation value of $ \braket{\fit/\nu} = 1 $, although there are subtleties associated with this interpretation here, as we will discuss, some of which have already been mentioned above.

\begin{table}[ht]
\centering
\footnotesize
\renewcommand{\arraystretch}{1.5}
\begin{tabular}{| c | r r | }
\hline
\rule{0pt}{4ex}\makecell{Parameter \\ {\scriptsize (from \cite{Gonzalez-Garcia:2014bfa})}}
						& bfp $ \pm1\sigma$ 			& 3$\sigma$ range \\[1.5ex] \hline\hline
\multicolumn{1}{| c |}{$ \sin^2 \theta_{12} $}	& 0.304$^{+0.013}_{-0.012}$		& 0.270 $\rightarrow$ 0.344	\\
$ \theta^l_{12} \,(^\circ)$			& 33.48 $^{+0.78}_{-0.75}$		& 31.29 $\rightarrow$ 35.91	\\ 
\multicolumn{1}{| c |}{$ \sin^2 \theta_{23} $}	& 0.452$^{+0.052}_{-0.028}$		& 0.382 $\rightarrow$ 0.643	\\ 
$ \theta^l_{23} \,(^\circ)$			& 42.3 $^{+3.0}_{-1.6}$		& 38.2 $\rightarrow$ 53.3 	\\ 
\multicolumn{1}{| c |}{$ \sin^2 \theta_{13} $}	& 0.0218$^{+0.0010}_{-0.0010}$	& 0.0186 $\rightarrow$ 0.0250	\\ 
$ \theta^l_{13} \,(^\circ)$			& 8.5 $^{+0.20}_{-0.21}$		& 7.85 $\rightarrow$ 9.10 	\\ 
$ \delta^l \,(^\circ)$				& 306 $^{+39}_{-70}$			& 0 $\rightarrow$ 360		\\ [5pt]
$\dfrac{\Delta m^2_{21}}{10^{-5}}$ eV$^2$	& 7.50 $^{+0.19}_{-0.17}$		& 7.02 $\rightarrow$ 8.09	\\ [8pt]
$\dfrac{\Delta m^2_{31}}{10^{-3}}$ eV$^2$ 	& +2.457 $^{+0.047}_{-0.047}$		& +2.317 $\rightarrow$ +2.607\\ [5pt]
\hline
\end{tabular}
\caption{
Table of current best fits to experimental data for neutrino mixing angles and masses in the case of normal mass squared ordering, taken from \cite{Gonzalez-Garcia:2014bfa}, with 1$\sigma$ and 3$\sigma$ uncertainty ranges. These are the values that we use in the CSD($n$) fits, apart from $ \delta_{\mathrm{CP}}$ which we leave as a prediction since its non-zero value has not yet been firmly established experimentally.}
\label{tab:data}
\end{table}

\subsection{Minimising method}
Initially, a coarse Monte-Carlo was used to examine the (5-dimensional) parameter space. A random vector $ x = (m_a,m_b,m_c,\eta,\xi) $ is chosen, all PMNS parameters are calculated numerically using the Mixing Parameter Tools (MPT) package for Mathematica \cite{Antusch:2009gu}, and $ \fit $ is evaluated. A large-$N$ search of this type reveals the existence of two collections of global minima. These regions in parameter space are characterised by having the same approximate values of $ m_a $ and $ m_b $, while $ m_c $ and $ \xi $ are allowed to take a broad range of values (in fact $ \xi $ can take any value at all in $ [-\pi,\pi] $). 

Meanwhile $ \eta $ is constrained only up to a sign -- the two minima then correspond to equal and opposite values of $ \eta $. Refining the input parameter space by allowing only $ \eta \in (0,\pi) $ leaves a single global minimum region. This minimum is well-defined and generally stable, meaning our $ \fit $ statistic is a good test for goodness-of-fit over this space; this is true for all CSD($n$). For more details on the behaviour of $ \fit $ near the global minimum, see Appendix \ref{sec:chisqinput}. Once the single global minimum is confirmed, numerical minimisation is performed in Mathematica by the method of differential evolution.

\FloatBarrier
\section{Results}
\label{sec:results}
This section details results for the properties of general CSD($n$) vacuum alignments, wherein we have simplified the analysis by considering only two planes of fixed $ \xi $, i.e. the cases where $ \xi = 0 $ (phase aligned with dominant mass matrix) or $ \xi = \eta $ (phase aligned with subdominant mass matrix). This simplification is predicated on the underlying assumption from CSD that the contribution from the $ m_c $ term in Eq.~\ref{eq:seesaw} is small; indeed, a stable minimum of the same order in $ \fit $ can be found for any value of $ \xi $. Such a constraint on $ \xi $ may also arise directly from a model, such as in \cite{King:2013hoa}. 

In all subsequent plots, a thick solid gridline corresponds to a best fit value of a mixing angle or neutrino mass, while thin solid gridlines show the $ 1\sigma $ limits, and thin dashed gridlines show the $ 3\sigma $ range.

\subsection{CSD(\secheadmath{n}) with two right-handed neutrinos}
Models with only two right-handed neutrinos are compelling as they are typically highly predictive. In a CSD($n$) framework, the neutrino mass matrix in Eq.~\ref{eq:mnu} simplifies in the two right-handed neutrino case to
\begin{equation}
	m^\nu_{(n)} = m_a \pmatr{0&0&0\\0&1&1\\0&1&1} + m_b e^{i\eta} \pmatr{1&n&n-2\\n&n^2&n(n-2)\\n-2&n(n-2)&(n-2)^2},
	\label{eq:mnu2}
\end{equation}
where we have defined $ \eta = \beta - \alpha $ and removed an overall unphysical phase $\alpha$. This case immediately predicts the lightest physical neutrino mass to be zero, $m_1=0$. For a given choice of alignment $n$, there are three real input parameters $ m_a $, $ m_b $ and $ \eta $ from which two light physical neutrino masses $ m_2 $, $ m_3 $, three lepton mixing angles, the CP-violating phase $ \deltacp $ and two Majorana phases are derived; a total of nine physical parameters from three input parameters, i.e. six predictions for each value of $n$. As the Majorana phases are not known and $ \deltacp $ is only tentatively constrained by experiment, this leaves five presently measured observables, namely the two neutrino mass squared differences and the three lepton mixing angles, from only three input parameters. 

\begin{table}[ht]
\renewcommand{\arraystretch}{1.2}
\centering
\footnotesize
\begin{tabular}{| c | c  c  c | c  c  c  c  c  c || c |}
\hline
\rule{0pt}{4ex} $n$ 	& \makecell{$m_a$ \\ {\scriptsize (meV)}} & \makecell{$m_b$ \\ {\scriptsize (meV)}} & 
\makecell{$\eta$  \\ {\scriptsize (rad)}}  	& \makecell{$\theta_{12}$ \\ {\scriptsize ($^{\circ}$)}} & \makecell{$\theta_{13}$ \\ {\scriptsize ($^{\circ}$)}}  & \makecell{$\theta_{23}$ \\ {\scriptsize ($^{\circ}$)}} & \makecell{$|\deltacp|$ \\ {\scriptsize ($^{\circ}$)}} & \makecell{$m_2$ \\ {\scriptsize (meV)}} & \makecell{$m_3$ \\ {\scriptsize (meV)}} & $\fit$  \\ [2ex] \hline 
1 	& 24.8		& 2.89		& 3.14		& 35.3		& 0 		& 45.0		& 0		& 8.66		& 49.6		& 485		\\ 
2 	& 19.7		& 3.66		& 0		& 34.5		& 7.65		& 56.0		& 0 		& 8.85		& 48.8		& 95.1		\\ 
3 	& 27.3		& 2.62		& 2.17		& 34.4		& 8.39		& 44.5		& 92.2		& 8.69		& 49.5		& 3.98		\\ 
4 	& 36.6		& 1.95		& 2.63		& 34.3		& 8.72		& 38.4		& 120		& 8.61		& 49.8		& 8.82		\\ 
5 	& 45.9		& 1.55		& 2.88		& 34.2		& 9.03		& 34.4		& 142		& 8.53		& 50.0		& 33.8		\\ 
6 	& 55.0		& 1.29		& 3.13		& 34.2		& 9.30		& 31.6		& 179		& 8.46		& 50.2		& 65.2		\\ 
7 	& 63.0		& 1.12		& 3.14		& 34.1		& 9.68		& 31.0		& 180		& 8.35		& 50.6		& 100		\\ 
8 	& 71.0		& 0.984		& 3.14		& 34.0		& 9.96		& 30.6		& 180		& 8.25		& 50.8		& 135		\\ 
9 	& 79.0		& 0.880		& 3.14		& 33.9		& 10.2		& 30.3		& 180		& 8.17		& 51.0		& 168		\\
\hline
\end{tabular}
\caption{Table of best fit parameters for two right-handed neutrino CSD($n$) model for $ 1 \leq n \leq 9 $. The fitted three input parameters  $ m_a $, $ m_b $ and $ \eta $ yield nine physical predictions, but only six physical outputs are shown. The undisplayed outputs are $m_1=0$ in each case and the two Majorana phases which are difficult to measure for a normal hierarchy.}
\label{tab:chisq2nu}
\end{table}

\begin{figure}[ht]
	\centering
	\includegraphics[scale=0.6]{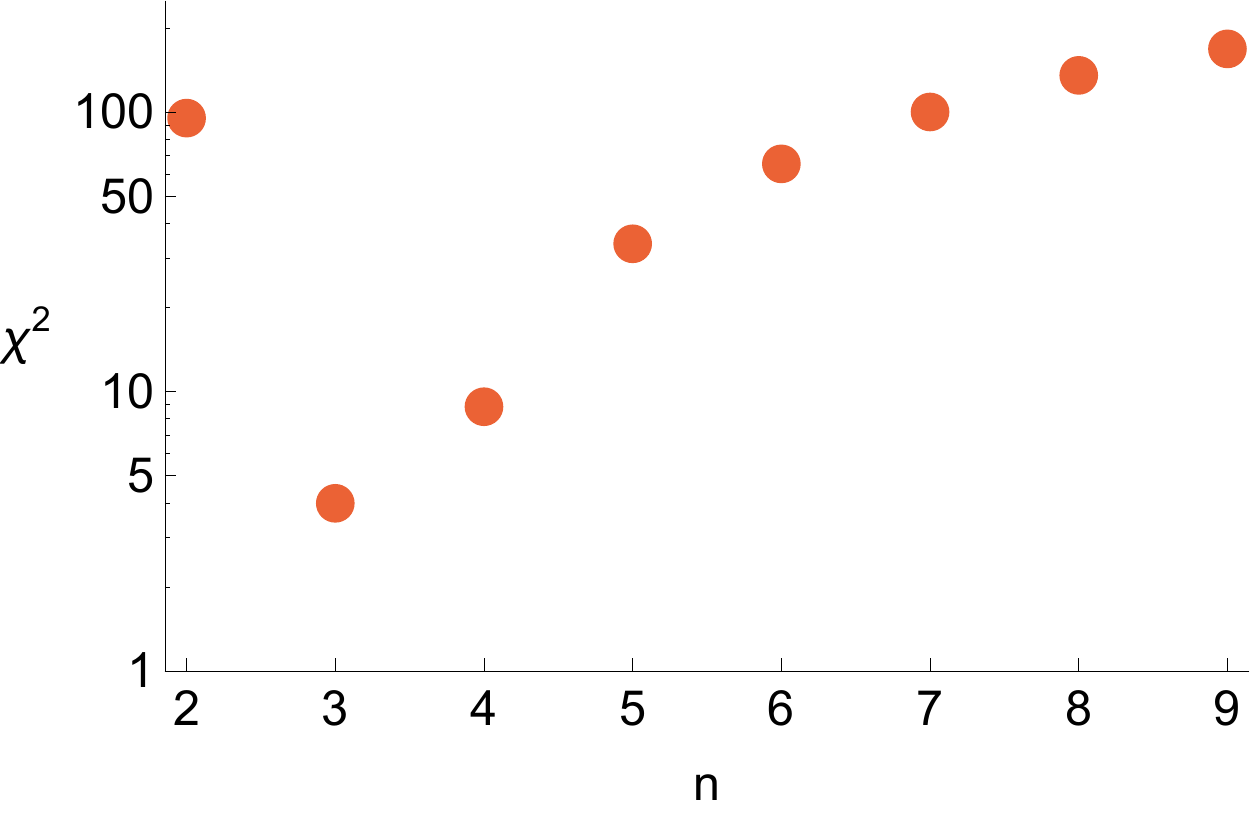}
	\caption{Best fit $ \fit $ with respect to $ n $.
	}
	\label{fig:chisq2nu}
\end{figure}

Table \ref{tab:chisq2nu} shows all fitted parameters with respect to $n$. \figref{fig:chisq2nu} shows the best fit values of $ \fit $ with respect to vacuum alignment $n$. Both CSD(3) and CSD(4) have $ \fit <10 $, while all others have significantly higher values, generally increasing with $ n $. With five values $ N $ fitted to three input parameters $ N_I $, this gives us two excess degrees of freedom, i.e. $ \nu \equiv N-N_I $ = 2. Recalling that in standard $ \fit $ analyses, $ \braket{\fit/\nu} = 1 $ for a good fit, for the most promising model, CSD(3), we have $ \chi^2/\nu =1.99 $.
We view this as a good fit,
particularly in light of the fact that it can naturally predict a CP phase $ \deltacp $ close to the current experimental preferred value of $ \sim -\pi/2 $. 
Similarly the fit for CSD(4) shows promise for model-building, with $ \braket{\fit/\nu} = 4.41 $ and a prediction $ |\deltacp| = 120^\circ $.
For $ n \geq 4 $, the largest contribution to $ \chi^2 $ is typically $ \theta_{23} $, while for $ n = 3 $ there is no dominant contribution.

In \figref{fig:mixingangles2nu} and \figref{fig:neutrinom2nu} we show the variation of physical masses and neutrino mixing angles with respect to $n$ in the two right-handed neutrino CSD($n$) model. Note that, in our conventions defined earlier, a positive value of $\eta$, namely $ \eta \in (0,\pi) $, yields a negative CP-violating angle, i.e.  $ \deltacp \in (0,-\pi) $, while the mirror global minimum for  $ \eta \in (-\pi,0) $ corresponds uniquely to  $ \deltacp \in (\pi,0) $. As $ \eta $ is unconstrained (unless some model explicitly restricts its domain), only the absolute value of $ \deltacp $ can be predicted by this analysis. In Table \ref{tab:chisq2nu} we only show positive $\eta$ values, for which $\deltacp$ is negative.

\begin{figure}[ht]
	\centering
	\includegraphics[scale=0.53]{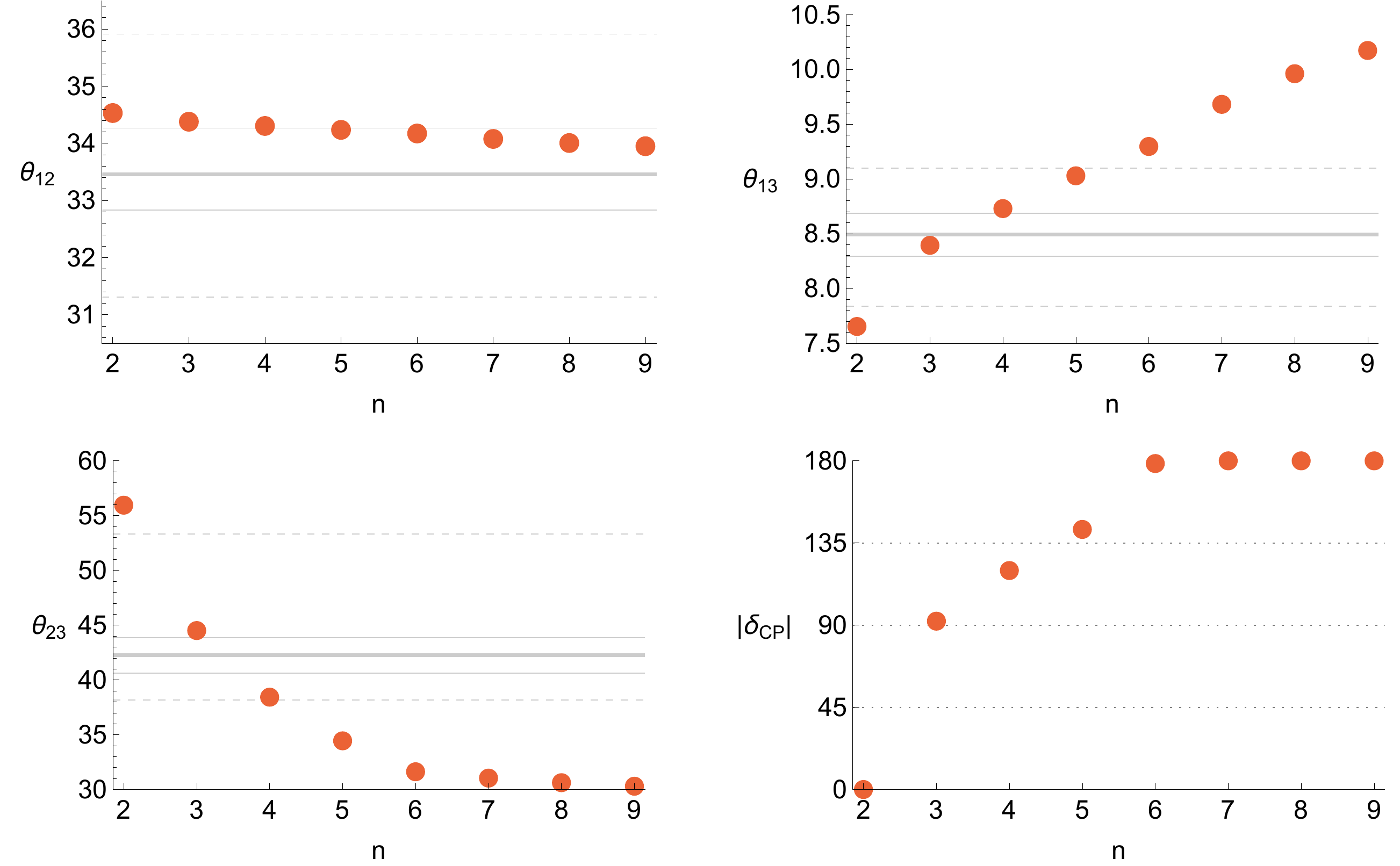}
	\caption{Best-fit PMNS mixing angles and CP-violating phase with respect to $n$, for the two right-handed neutrino CSD($n$) model. We emphasise that $|\delta_{\rm CP} |$ is a genuine prediction here since have not used the one sigma hint from experiment as an input constraint. It is striking that both CSD(3) and CSD(4) both yield predictions within the preferred range $|\delta_{\rm CP} |\sim 90^{\circ}\pm 45^{\circ}$ but may be distinguished by their differing predictions for the atmospheric angle $\theta_{23}\approx 45^{\circ}$ and $\theta_{23}\approx 38^{\circ}$, respectively.}
	\label{fig:mixingangles2nu}
\end{figure}

\begin{figure}[hb]
	\centering
	\includegraphics[scale=0.53]{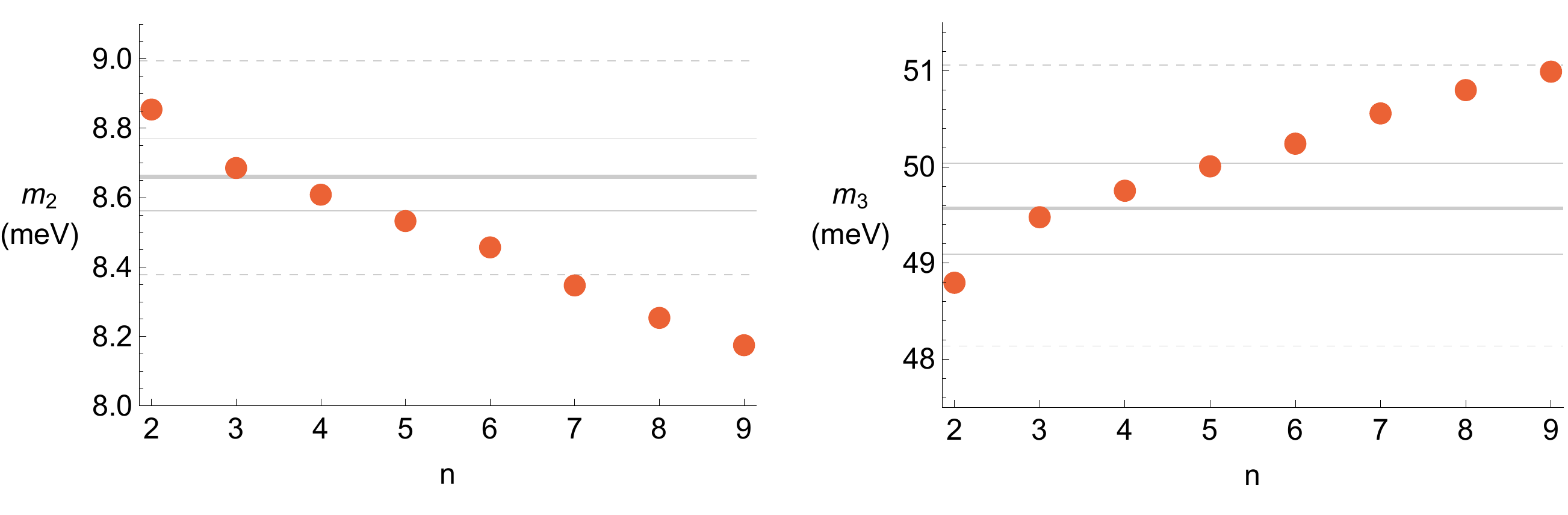}
	\caption{Best-fit light neutrino masses with respect $n$, for the two right-handed neutrino CSD($n$) model. Since $ m_1 =0 $ in this case $ m_2 = \sqrt{\Delta m_{21}^2} $ and $ m_3 = \sqrt{\Delta m_{31}^2} $.}
	\label{fig:neutrinom2nu}
\end{figure}

As discussed in Appendix~\ref{sec:chisqinput}, the two input masses $m_a$, $m_b$ fix the two light neutrino masses $ m_2 $, $ m_3 $ after which the entire PMNS matrix is determined from only one parameter, namely the phase $\eta$. {\it A priori}, CSD($n$) need not lead to low $\fit$ values for any choice of $n$, due to the sensitivity of the predictions to the phase $\eta$, yet in fact the results show that it gives very good fits to the leptonic mixing angles for $n=3,4$, for special values of $\eta$, yielding a value of $|\delta_{\mathrm{CP}}|$ (which is taken to be unconstrained by data) as a genuine prediction, along with preferred values for the lepton angles.

This is illustrated in \figref{fig:2nuchisq} which shows the variation of $ \fit $ with $ \eta $, for CSD($n$) with $ 1 \leq n \leq 9 $. It is clear that $ \eta $ is quite strongly constrained, even for CSD(3) and CSD(4), which can give good fits; with CSD(3), the values (in radians) of $ \eta $ that give $ \fit<10 $ are $ 2.08 \simlt \eta \simlt 2.27 $, which is a range of approximately $ 11^\circ $. This range happens to include the value $ 2\pi/3 $. Such a value could be produced in a model with a discrete symmetry such as $ \mathbb{Z}_{3N} $.

The neutrino masses are also tightly constrained. Recalling the best fit values given in Table \ref{tab:chisq2nu}, any fit that yields $ \fit \simlt 50 $ will correspond to values of $ m_a $ and $ m_b $ that are within $ \pm $10-15\% of their best fit value. This is true for all CSD($n$). In other words, the ranges of acceptable values for the input masses scale with the best fit value. This is also confirmed for models with three right-handed neutrinos, discussed below, and is discussed further in Appendix \ref{sec:chisqinput}.

\begin{figure}[ht]
	\centering
	\includegraphics[scale=0.4]{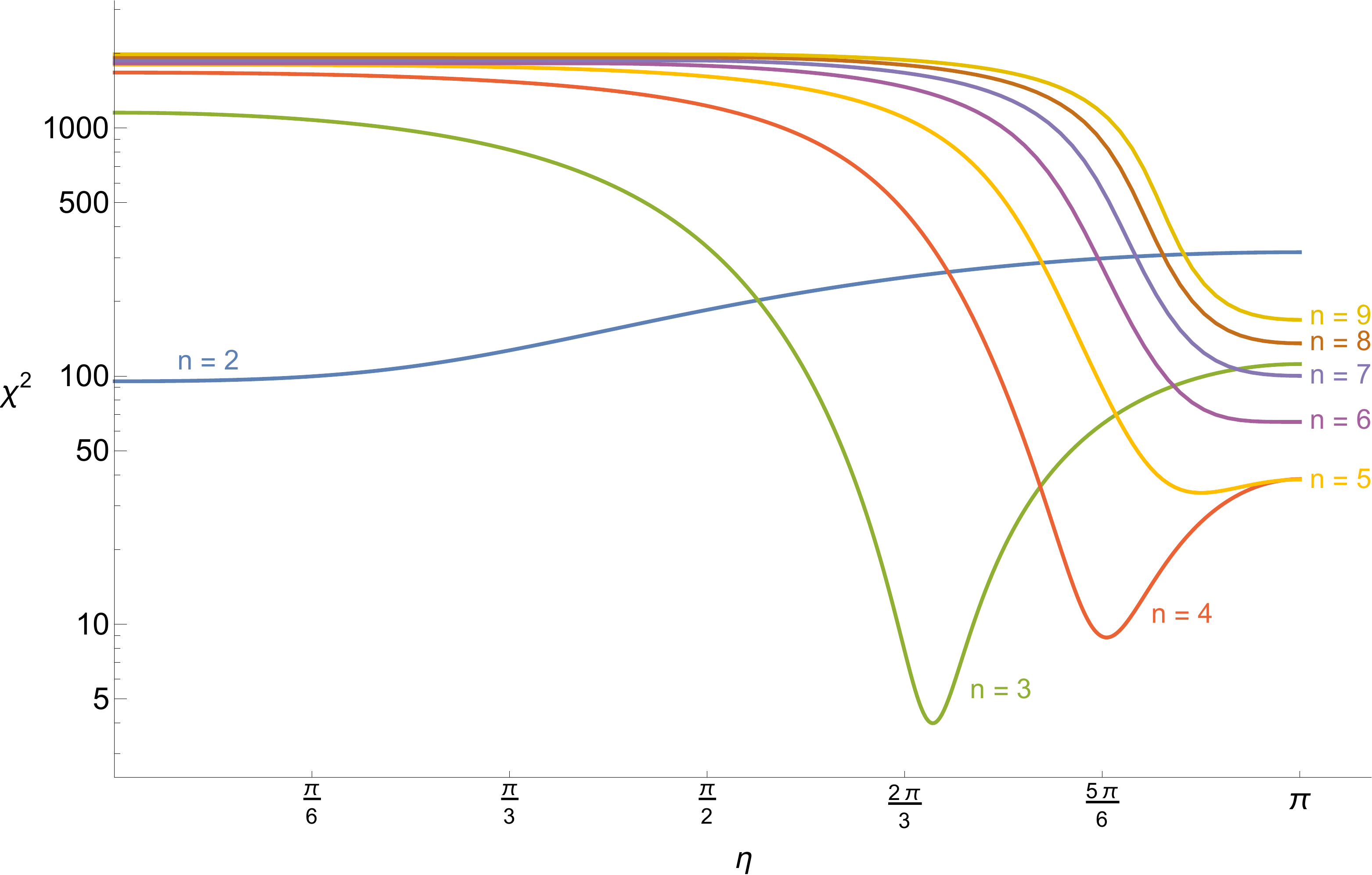}
	\caption{Variation of $ \fit $ with phase $ \eta $ in a model with two right-handed neutrinos. 
	}
\label{fig:2nuchisq}
\end{figure}

To make the link between $ \fit $ minimisation and physical prediction more concrete, we examine in \figref{fig:mixingangles_eta_csd3-5} the variation in the three mixing angles with $ \eta $, for the physically most interesting cases of CSD($ n $) with $ n = 3,4,5 $. We see that although $ \theta_{12} $ is largely unaffected by $ \eta $, there is a complicated dependence of the other two mixing angles on $ \eta $, which is different for different $ n $. These plots demonstrate what the $ \fit $ value suggests: for some small set of values $ \eta $, the predicted mixing angles converge on the experimental best fit values for CSD(3) and CSD(4). Meanwhile for CSD(5) we begin to see tension between the fits to $ \theta_{13} $ and $ \theta_{23} $; this tension grows with large $ n $.

\begin{figure}[ht]
	\centering
	\begin{subfigure}{0.95\textwidth}
		\centering
		\includegraphics[scale=0.59]{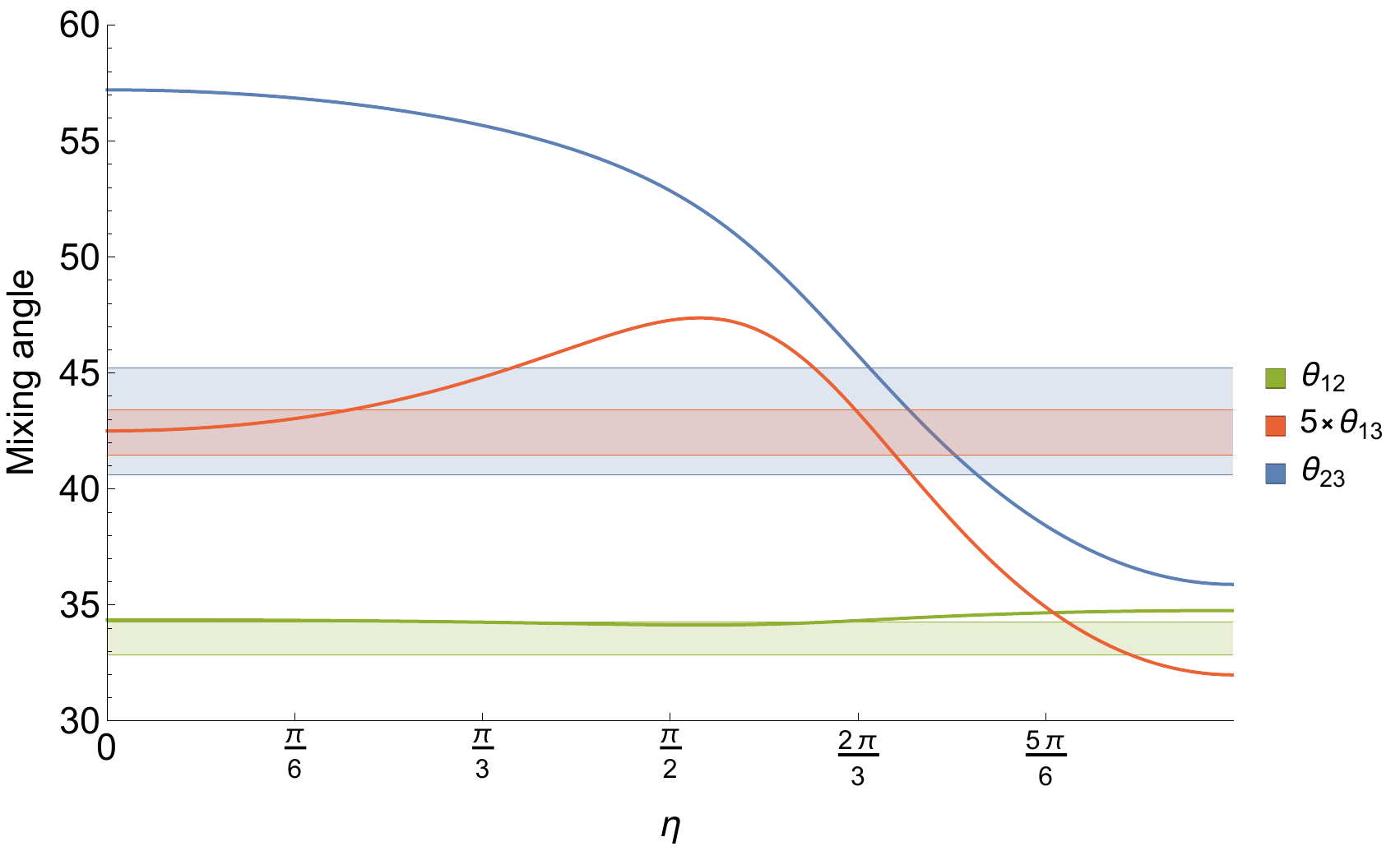}
		\caption{CSD(3)}
	\end{subfigure} 
	\begin{subfigure}{0.95\textwidth}
		\centering
		\includegraphics[scale=0.59]{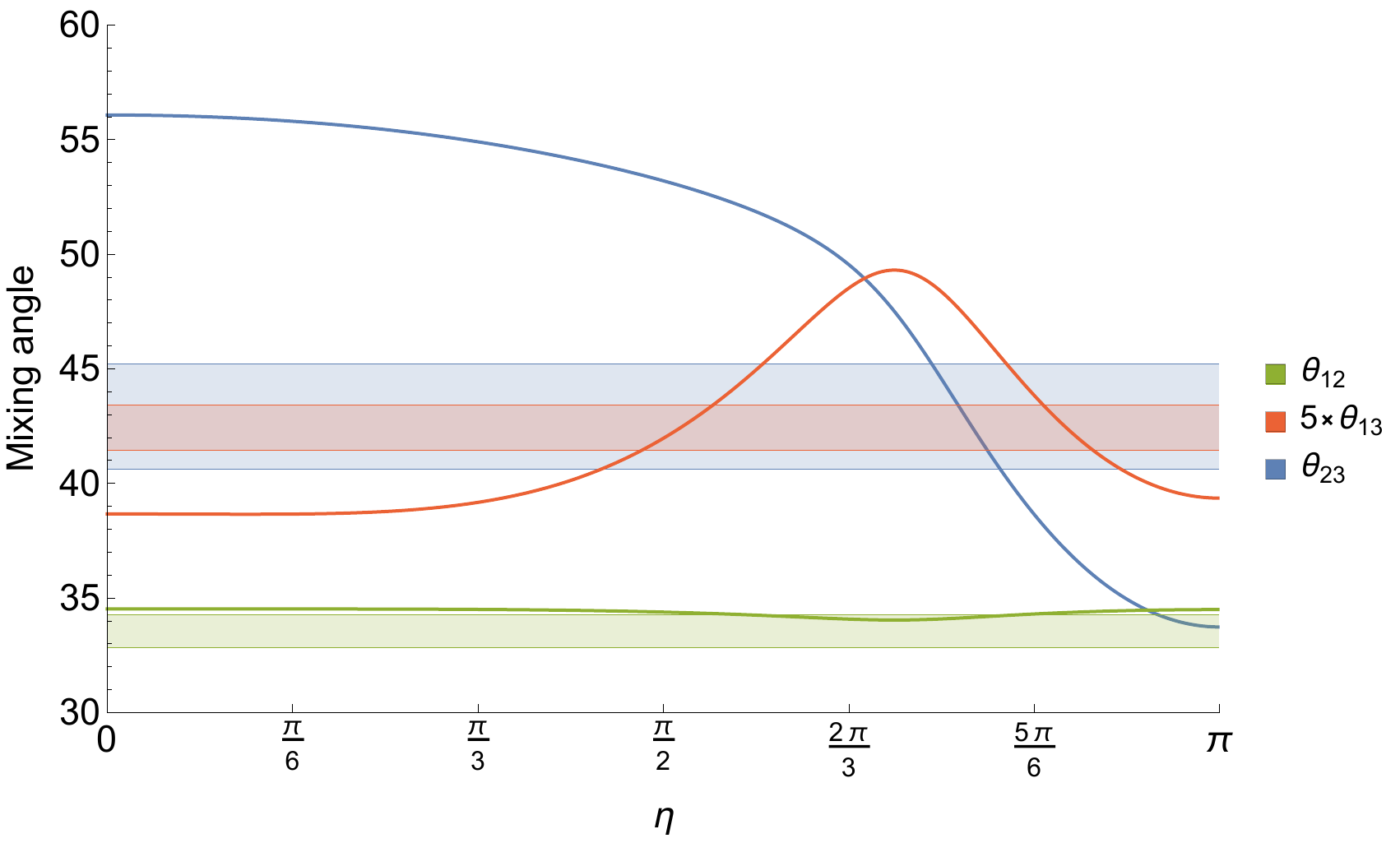}
		\caption{CSD(4)}
	\end{subfigure} 
	\begin{subfigure}{0.95\textwidth}
		\centering
		\includegraphics[scale=0.59]{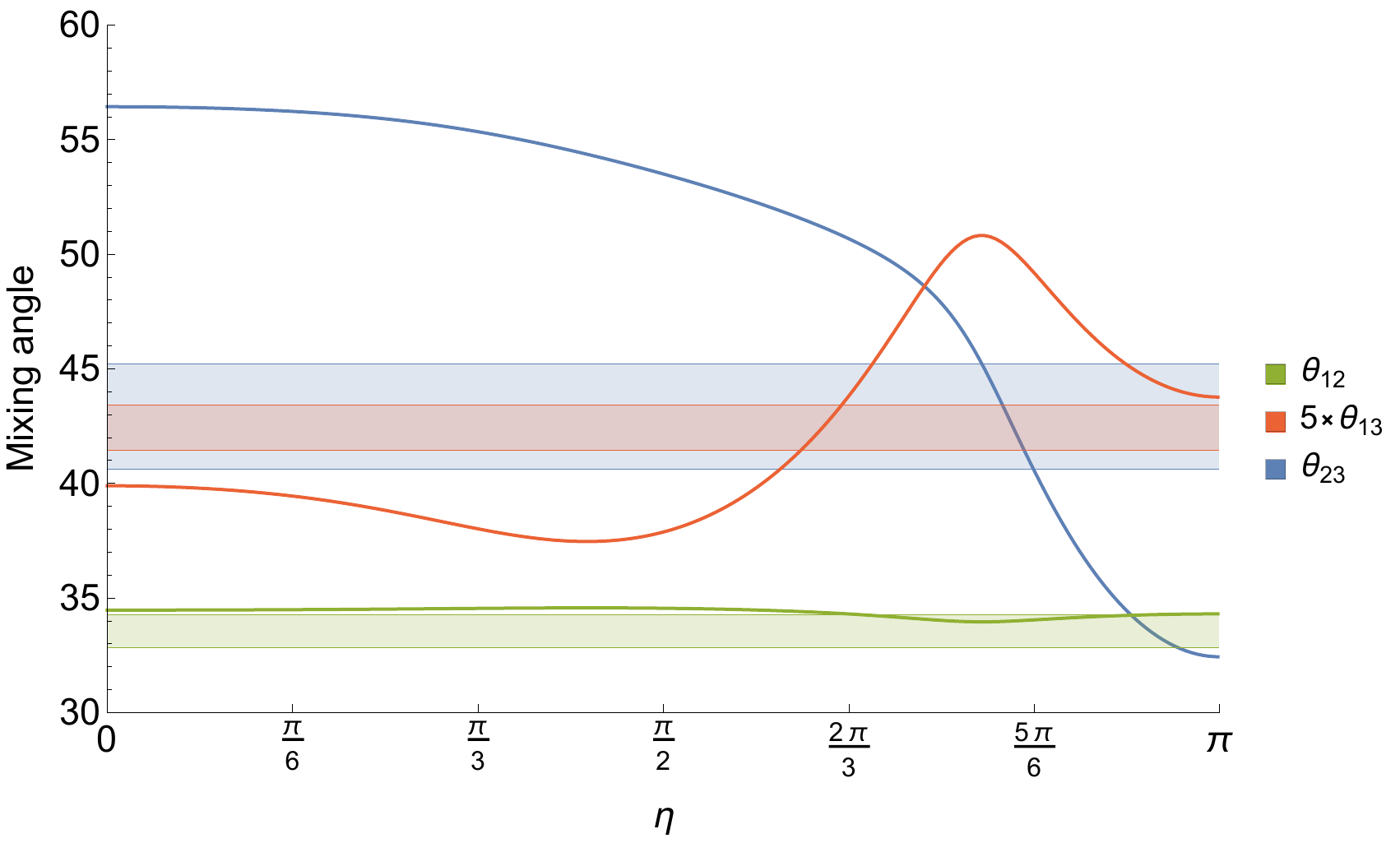}
		\caption{CSD(5)}
	\end{subfigure}
	\caption{Variation of the three lepton mixing angles with phase $ \eta $ in models with two right-handed neutrinos. Shaded regions represent the $ \pm1\sigma $ range for each mixing angle (in colours corresponding to the drawn curve for $ \theta_{ij} $). The reactor angle $ \theta_{13} $ has been multiplied by a factor 5 for the sake of visual ease (there is no physical relevance of the number 5).}
	\label{fig:mixingangles_eta_csd3-5}
\end{figure}

\FloatBarrier
\subsection{CSD(\secheadmath{n}) with three right-handed neutrinos}
We now extend the analysis to the case of three right-handed neutrinos,
\begin{equation}
	m^\nu_{(n)} = m_a \pmatr{0&0&0\\0&1&1\\0&1&1} + m_b e^{i\eta} \pmatr{1&n&n-2\\n&n^2&n(n-2)\\n-2&n(n-2)&(n-2)^2} + m_c e^{i\xi} \pmatr{0&0&0\\0&0&0\\0&0&1},
	\label{eq:mnu3}
\end{equation}
where the overall unphysical phase $\alpha$ in Eq.~\ref{eq:mnu} has been factored out and dropped. The immediate effect of including a third right-handed neutrino is to switch on a non-zero value for the lightest physical neutrino mass $m_1$, where previously for the case of two right-handed neutrinos we had $m_1=0$.

Since the contribution from the ``decoupled'' right-handed neutrino is assumed to be a perturbation to the case of two right-handed neutrinos considered in the previous subsection, the detailed structure of the third matrix is irrelevant, and it is sufficient to only keep the most important term in the third matrix, which we have assumed to be in the (3,3) entry, since in unified models where $Y^u=Y^{\nu}$ this entry is responsible for the top quark Yukawa coupling \cite{King:2013hoa}. However the third term brings in a further undetermined relative phase $\xi$ which complicates the analysis somewhat. Since the results are comparatively less sensitive to this phase $\xi$, particularly for the physically interesting cases of $ n = 3,4 $, we have taken the approach of fixing it to take two simple values, namely  $ \xi = 0 $ and $ \xi = \eta $, in order to illustrate the sensitivity of the results to this phase without over-complicating the analysis. These two examples also correspond to the values appearing in certain realistic models \cite{King:2013hoa}. 

Once the existence of a single stable minimum has been confirmed, Tables  \ref{tab:chisq_xi0} and \ref{tab:chisq} show the results for the best fit or optimal $ \fit $ and its corresponding input and output values, for CSD($n$) with $ 1 \leq n \leq 9 $, for each of the two sub-subdominant phases considered, i.e. $ \xi = 0 $ and $ \xi = \eta $. As in the two right-handed neutrino case, CSD(3) and CSD(4) can achieve $ \fit < 10 $. More generally for each CSD($n$), the raw $ \fit $ values are slight improvements over the two neutrino case, which is expected as we have added a free parameter $ m_c $. 

Evaluating the number of excess degrees of freedom $ \nu = N - N_I $ is non-trivial. We have added two input parameters $ m_c $ and $ \xi $, giving nominally $ N_I = 5 $ and hence $\nu =0$. However in practice we have fixed $\xi$ to some convenient values (since the fit is insensitive to $\xi$) so this gives $ N_I = 4$ and hence $\nu =1$. Moreover, $m_c$ is constrained by the SD assumption that the third right-handed neutrino is nearly decoupled from the theory. Therefore it may be more reasonable to compare the expected $\chi^2$ value to some finite value rather than zero, but the precise value is debatable. One may cautiously regard $\chi^2$ values between unity and, say, up to 10 as encouraging, bearing in mind also that we do not include $\deltacp$ in the fit, and also that we are conservative with the asymmetric error in the atmospheric angle, as discussed earlier. On the other hand the $n$ itself in CSD($n$) may be regarded as a further discrete parameter. In the light of all of the above, we should interpret $\fit$ values with care.

\begin{table}[ht]
\renewcommand{\arraystretch}{1.2}
\centering
\footnotesize
\begin{tabular}{| c | c  c  c  c | c  c  c  c  c  c  c || c |}
\hline
\rule{0pt}{4ex} $n$ 	& \makecell{$m_a$ \\ {\scriptsize (meV)}} & \makecell{$m_b$ \\ {\scriptsize (meV)}} & \makecell{$m_c$ \\ {\scriptsize (meV)}} & \makecell{$\eta$  \\ {\scriptsize (rad)}} 
& \makecell{$\theta_{12}$ \\ {\scriptsize ($^{\circ}$)}} & \makecell{$\theta_{13}$ \\ {\scriptsize ($^{\circ}$)}}  & \makecell{$\theta_{23}$ \\ {\scriptsize ($^{\circ}$)}} & \makecell{$|\deltacp|$ \\ {\scriptsize ($^{\circ}$)}} & \makecell{$m_1$ \\ {\scriptsize (meV)}} & \makecell{$m_2$ \\ {\scriptsize (meV)}} & \makecell{$m_3$ \\ {\scriptsize (meV)}} & $\fit$ 	\\ [2ex] \hline 
1 	& 23.3		& 2.81		& 5.77		& 1.62		& 33.5		& 0.293		& 41.4		& 245		& 0.874		& 8.71		& 49.6		& 474		\\ 
2 	& 19.7		& 3.66		& 0		& 0		& 34.5		& 7.65		& 56.0		& 0 		& 0		& 8.85		& 48.8		& 95.1		\\ 
3 	& 26.0		& 2.60		& 1.77		& 2.1		& 33.6		& 8.37		& 44.6		& 81.3		& 0.278		& 8.69		& 49.5		& 2.59		\\ 
4 	& 32.3		& 1.94		& 4.75		& 2.48		& 33.0		& 8.70		& 38.8		& 89.1		& 0.692		& 8.64		& 49.7		& 6.51		\\ 
5 	& 38.3		& 1.52		& 7.10		& 2.65		& 32.4		& 8.92		& 35.6		& 89.2		& 0.964		& 8.62		& 49.9		& 25.1		\\ 
6 	& 44.5		& 1.25		& 9.81		& 2.74		& 31.8		& 9.04		& 33.6		& 88.6		& 1.12		& 8.61		& 50.0		& 43.1		\\ 
7 	& 50.7		& 1.06		& 10		& 2.81		& 31.3		& 9.12		& 32.3		& 87.9		& 1.22		& 8.61		& 50.1		& 58.1		\\ 
8 	& 57.3		& 0.92		& 10		& 2.85		& 31.0		& 9.29		& 32.0		& 87.5		& 1.23		& 8.57		& 50.1		& 70.9		\\ 
9 	& 64.0		& 0.82		& 10		& 2.88		& 30.7		& 9.44		& 32.1		& 86.9		& 1.22		& 8.54		& 50.2		& 82.4		\\
\hline
\end{tabular}
\caption{Table of best fit parameters for CSD($n$) for $ 1 \leq n \leq 9 $ and $ \xi = 0 $.}
\label{tab:chisq_xi0}
\end{table}

\begin{table}[ht]
\renewcommand{\arraystretch}{1.2}
\centering
\footnotesize
\begin{tabular}{| c | c  c  c  c | c  c  c  c  c  c  c || c |}
\hline
\rule{0pt}{4ex} $n$ 	& \makecell{$m_a$ \\ {\scriptsize (meV)}} & \makecell{$m_b$ \\ {\scriptsize (meV)}} & \makecell{$m_c$ \\ {\scriptsize (meV)}} & 
\makecell{$\eta$  \\ {\scriptsize (rad)}} 
	& \makecell{$\theta_{12}$ \\ {\scriptsize ($^{\circ}$)}} & \makecell{$\theta_{13}$ \\ {\scriptsize ($^{\circ}$)}}  & \makecell{$\theta_{23}$ \\ {\scriptsize ($^{\circ}$)}} & \makecell{$|\deltacp|$ \\ {\scriptsize ($^{\circ}$)}} & \makecell{$m_1$ \\ {\scriptsize (meV)}} & \makecell{$m_2$ \\ {\scriptsize (meV)}} & \makecell{$m_3$ \\ {\scriptsize (meV)}} & $\fit$  \\ [2ex] \hline 
1 	& 24.5		& 2.75		& 1.26		& 0		& 33.3		& 0.069		& 44.2		& 180		& 0.197		& 8.66		& 49.6		& 477		\\ 
2 	& 19.7		& 3.66		& 0		& 0		& 34.5		& 7.65		& 56.0		& 0		& 0		& 8.85		& 48.8		& 95.1		\\ 
3 	& 27.3		& 2.61		& 0.558		& 2.16		& 33.7		& 8.37		& 44.8		& 92.7		& 0.092		& 8.69		& 49.5		& 3.14		\\ 
4 	& 36.8		& 1.93		& 1.30		& 2.63		& 33.0		& 8.67		& 39.0		& 123		& 0.215		& 8.62		& 49.7		& 5.53		\\ 
5 	& 46.5		& 1.52		& 1.85		& 2.91		& 32.5		& 8.93		& 35.2		& 149		& 0.307		& 8.55		& 50.0		& 27.6		\\ 
6 	& 55.4		& 1.27		& 2.15		& 3.14		& 32.1		& 9.27		& 33.1		& 180		& 0.356		& 8.46		& 50.2		& 56.8		\\ 
7 	& 63.4		& 1.10		& 2.2		& 3.14		& 32.0		& 9.66		& 32.6		& 180		& 0.364		& 8.34		& 50.6		& 92.4		\\ 
8 	& 71.4		& 0.97		& 2.16		& 3.14		& 32.0		& 9.95		& 32.1		& 180		& 0.358		& 8.24		& 50.9		& 129		\\ 
9 	& 79.3		& 0.87		& 2.05		& 3.14		& 32.0		& 10.2		& 31.7		& 180		& 0.341		& 8.15		& 51.1		& 163		\\
\hline
\end{tabular}
\caption{Table of best fit parameters for CSD($n$) for $ 1 \leq n \leq 9 $ and $ \xi = \eta $.}
\label{tab:chisq}
\end{table}

As $n$ increases, the global fit prefers a stronger hierarchy of input neutrino masses $ m_a $ and $ m_b $, while the contribution from $ m_c $ becomes stronger. The fits select the input mass parameters $m_a$, $m_b$ and $m_c$ which are allowed to be free apart from an upper limit imposed on $m_c<10 $ meV in order not to violate the condition of sequential dominance as discussed in the Introduction. In the case of $ \xi = 0 $, $ m_c $ reaches the soft upper bound of 10 meV for CSD($ n \geq 7 $).%
\footnote{{Note that a fit that requires a large $ m_c $ is not CSD. For example, a proper analysis of such a non-CSD model necessarily includes contributions from elements of the sub-subdominant mass matrix other than the largest (3,3) element, which have been neglected thus far. This would destroy the predictivity of the scheme which makes CSD($n$) so appealing. This justifies imposing the chosen upper bound on $ m_c $.}}
However for the successful cases CSD(3) and CSD(4), the best fit value of $m_c$ are comfortably below 10 meV, so these cases naturally prefer a quite decoupled third right-handed neutrino for which the upper limit of $m_c$ is irrelevant.
Consequently, restricting our analysis to only examine two values of $ \xi $ is justified for small $ n $. For larger $ n $, the overall contribution from the third matrix proportional to $ m_c e^{i\xi} $ is larger, yet nevertheless fails to significantly improve the (poor) fit to data.

Among physical parameters, of particular note is the CP-violating phase $ |\deltacp| $, which is close to 90$^\circ$ for CSD(3) for both choices of $ \xi $. Furthermore, the alignment of $ \xi $ with the dominant or subdominant mass matrix appears to greatly affect the prediction for $ \deltacp $. The most likely source of this behaviour is a close relationship between $ \eta $ and $ \deltacp $. Notice that when $ \xi = \eta $, the best fit of both is 180$^\circ$ for $ n \geq 6 $. 
The observed plateaus in $ \deltacp $ are believed to be physical rather than artefacts of numerical minimisation, though an analytic treatment would be required for a deeper understanding of their origin. Note, however, that this behaviour only appears for CSD($n\geq 6$) with poor fits.

In \figref{fig:mixingangles_n_csd2-9} and \figref{fig:neutrinom_n_csd2-9} we display the variation of the best fit physical parameters -- mixing angles and neutrino masses -- as a function of $n$. In \figref{fig:mixingangles_n_csd2-9} we see that the reactor angle increases with $n$ while the atmospheric and solar angles decrease. Examining the $ 3\sigma $ ranges (dashed lines) we also see that $ \theta_{23} $ is typically worst fit (only CSD(3) lies within the $ 1\sigma $ bounds), and is also least sensitive to the choice of sub-subdominant phase $ \xi $, which can otherwise improve the fit of $ \theta_{12} $ or $ \theta_{13} $ at large $n$. 

We note the similarities between the mixing angle predictions in \figref{fig:mixingangles_n_csd2-9}-\ref{fig:neutrinom_n_csd2-9} and the matching figures for models with two neutrinos (\figref{fig:mixingangles2nu}-\ref{fig:neutrinom2nu}). The primary difference when a third neutrino is introduced is that $ \theta_{12} $ is pushed to a lower value. The best fit values of $ m_1 $ in \figref{fig:neutrinom_n_csd2-9} indicates that it can vary greatly with $n$ for some phase choices. It is however unlikely that this can be used to constrain models in the near future, as the mass scale is well below current experimental bounds of $ \sum m_\nu < 0.23 $ eV \cite{Ade:2013zuv}. 

\begin{figure}[ht]
	\centering
	\includegraphics[scale=0.75]{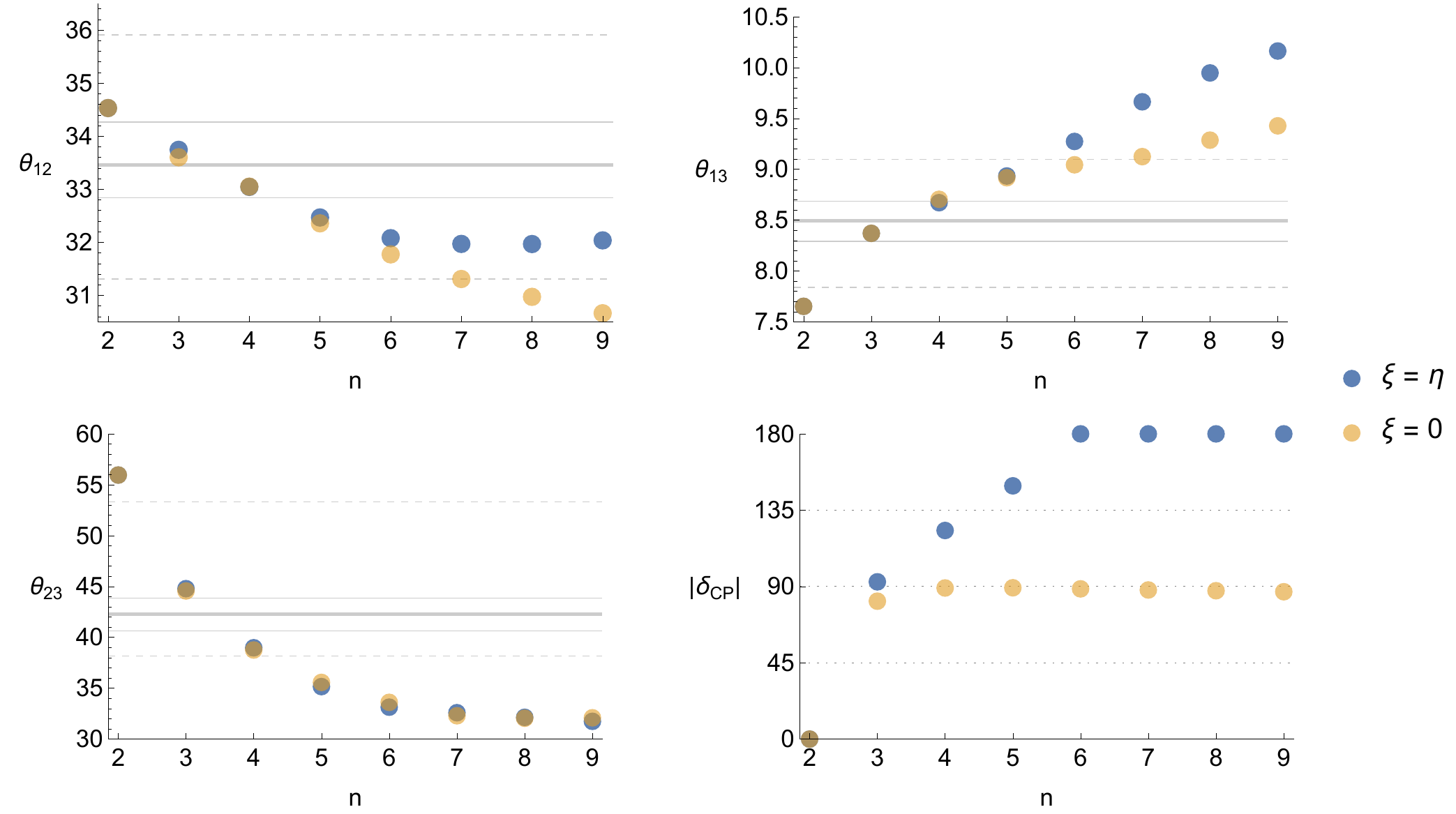}
	\caption{Best-fit PMNS mixing angles and the CP-violating phase $\deltacp$
	for CSD($n$) with three right-handed neutrinos as a function of $n$.
	The cases $\xi=0$ ($\xi=\eta$) correspond to yellow (blue) dots.}
	\label{fig:mixingangles_n_csd2-9}
\end{figure}

\begin{figure}[ht]
	\centering
	\includegraphics[scale=0.75]{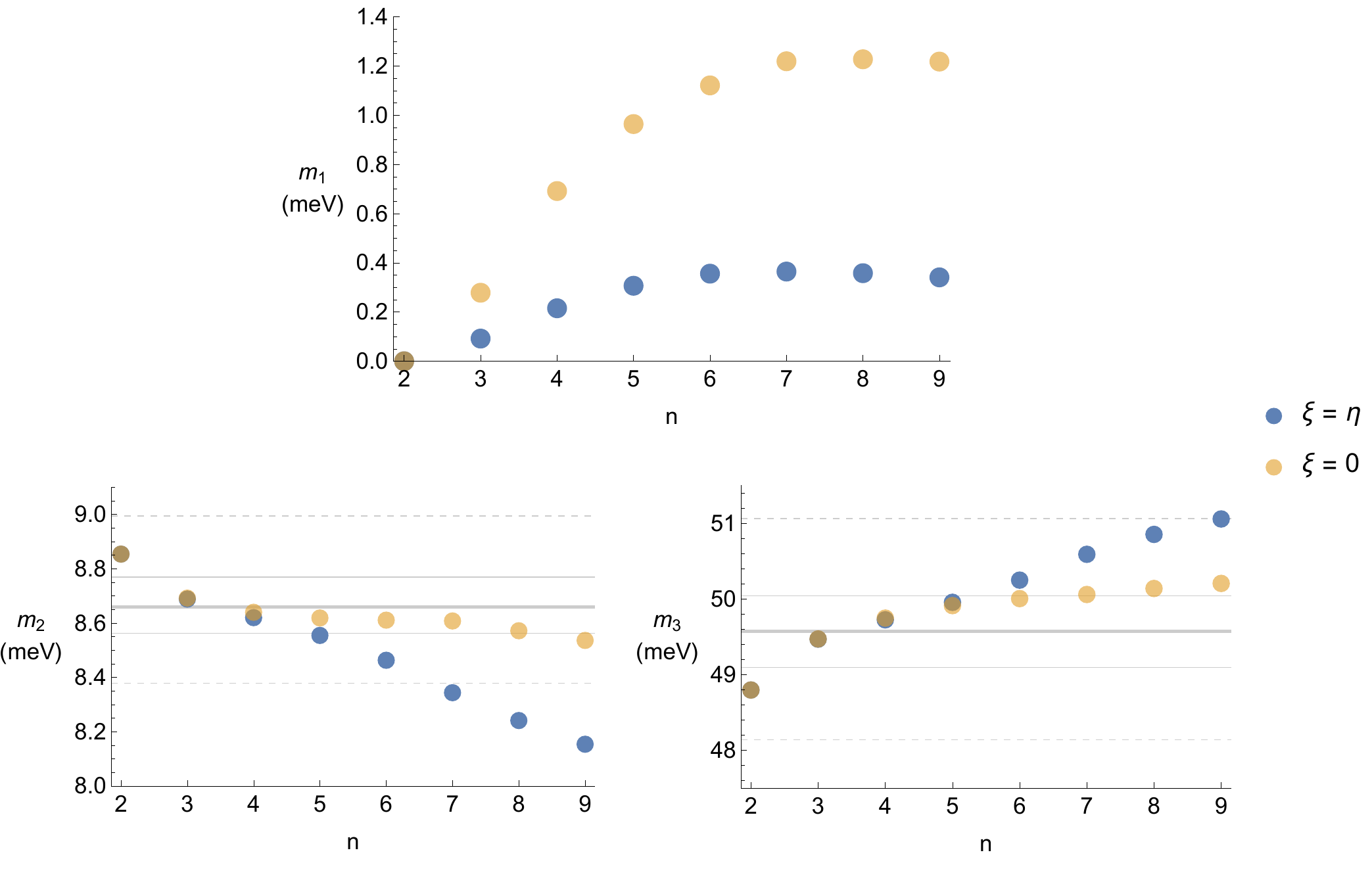}
	\caption{Best-fit light neutrino masses for CSD($n$) with three right-handed neutrinos as a function of $n$. The horizontal lines drawn assume $ m_1 $ is negligible, such that $ m_2 \simeq \sqrt{\Delta m_{21}^2} $ and $ m_3 \simeq \sqrt{\Delta m_{31}^2} $. The cases $\xi=0$ ($\xi=\eta$) correspond to yellow (blue) dots.}
	\label{fig:neutrinom_n_csd2-9}
\end{figure}

\begin{figure}[ht]
	\centering
	\begin{subfigure}{0.50\textwidth}
		\centering
		\includegraphics[scale=0.46]{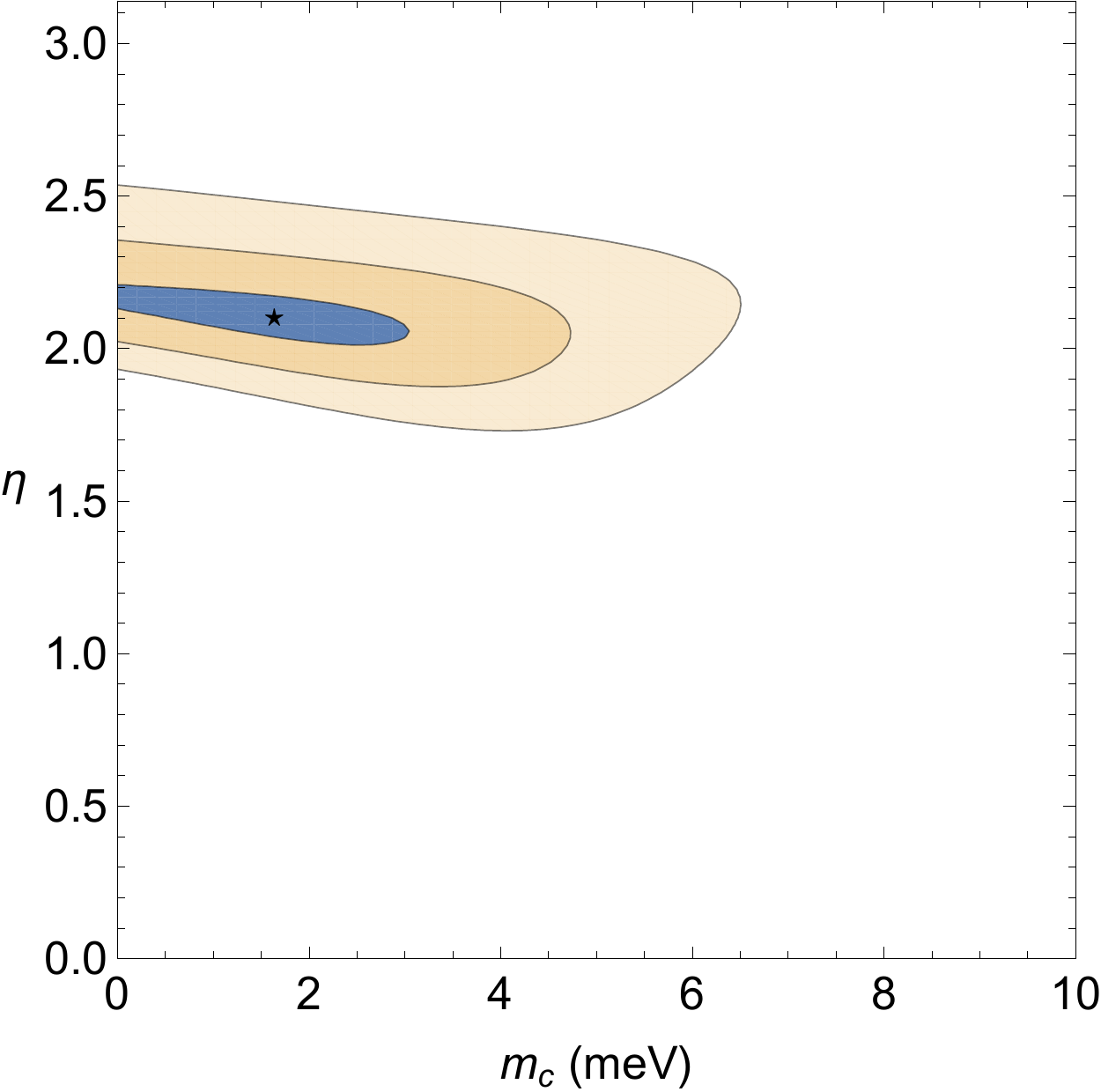}
		\caption{CSD(3), $ \xi = 0 $}
	\end{subfigure}%
	\hspace*{-2em}
	\begin{subfigure}{0.50\textwidth}
		\centering
		\includegraphics[scale=0.46]{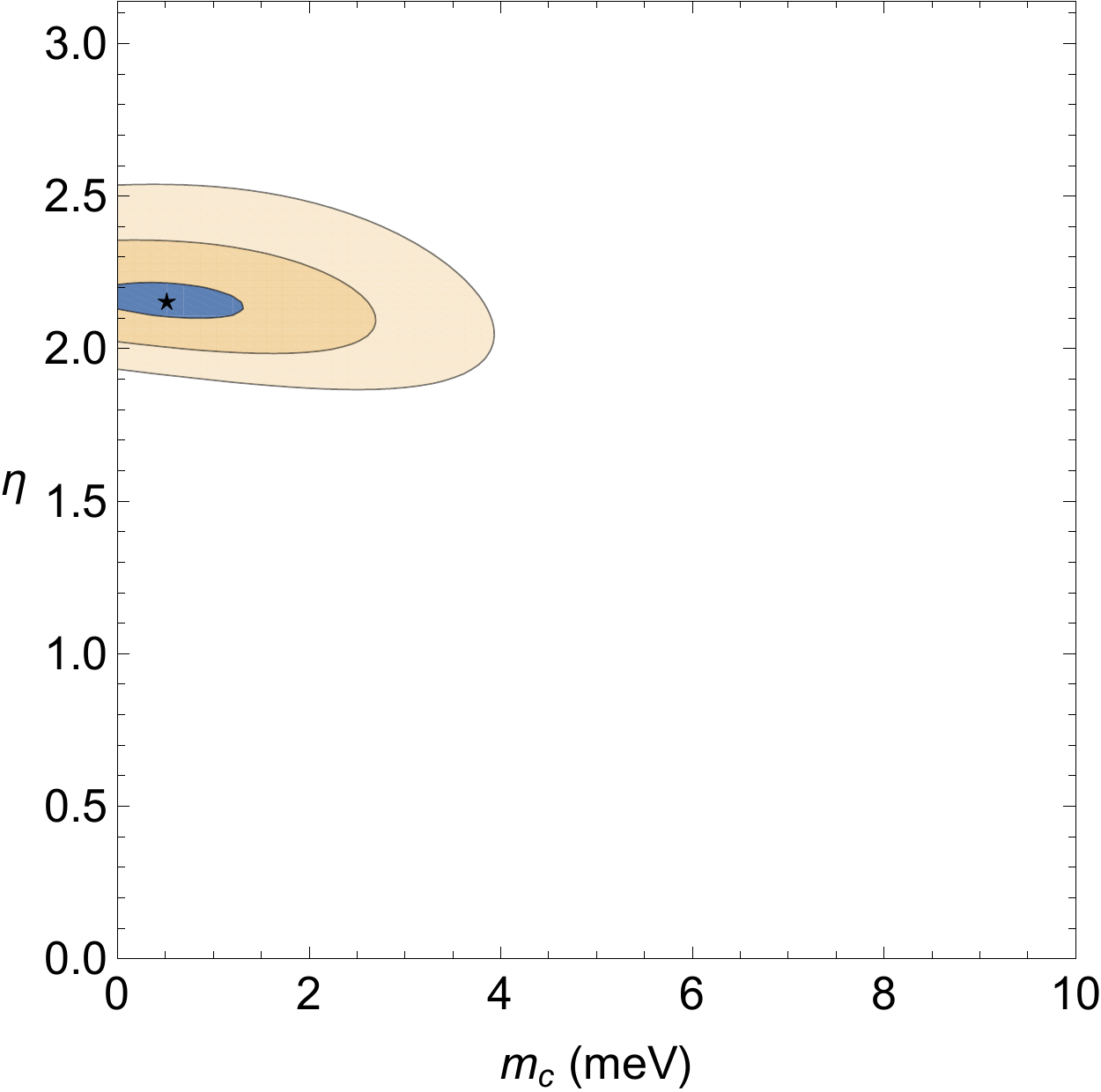}
		\caption{CSD(3), $ \xi = \eta $}
	\end{subfigure}\vspace{0.5cm}
	\begin{subfigure}{0.50\textwidth}
		\centering
		\includegraphics[scale=0.46]{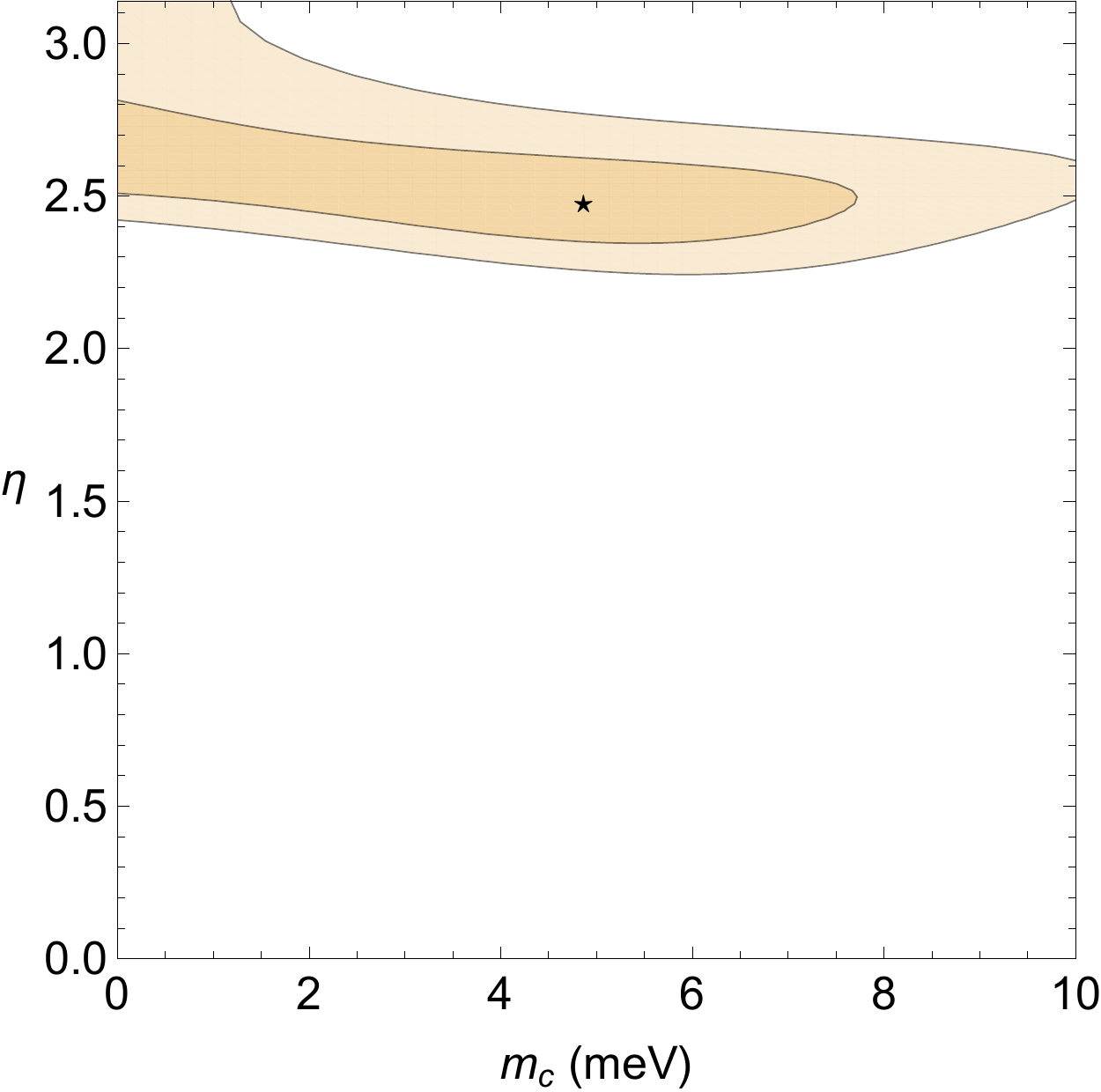}
		\caption{CSD(4), $ \xi = 0 $}
	\end{subfigure}%
	\hspace*{-2em}
	\begin{subfigure}{0.50\textwidth}
		\centering
		\includegraphics[scale=0.46]{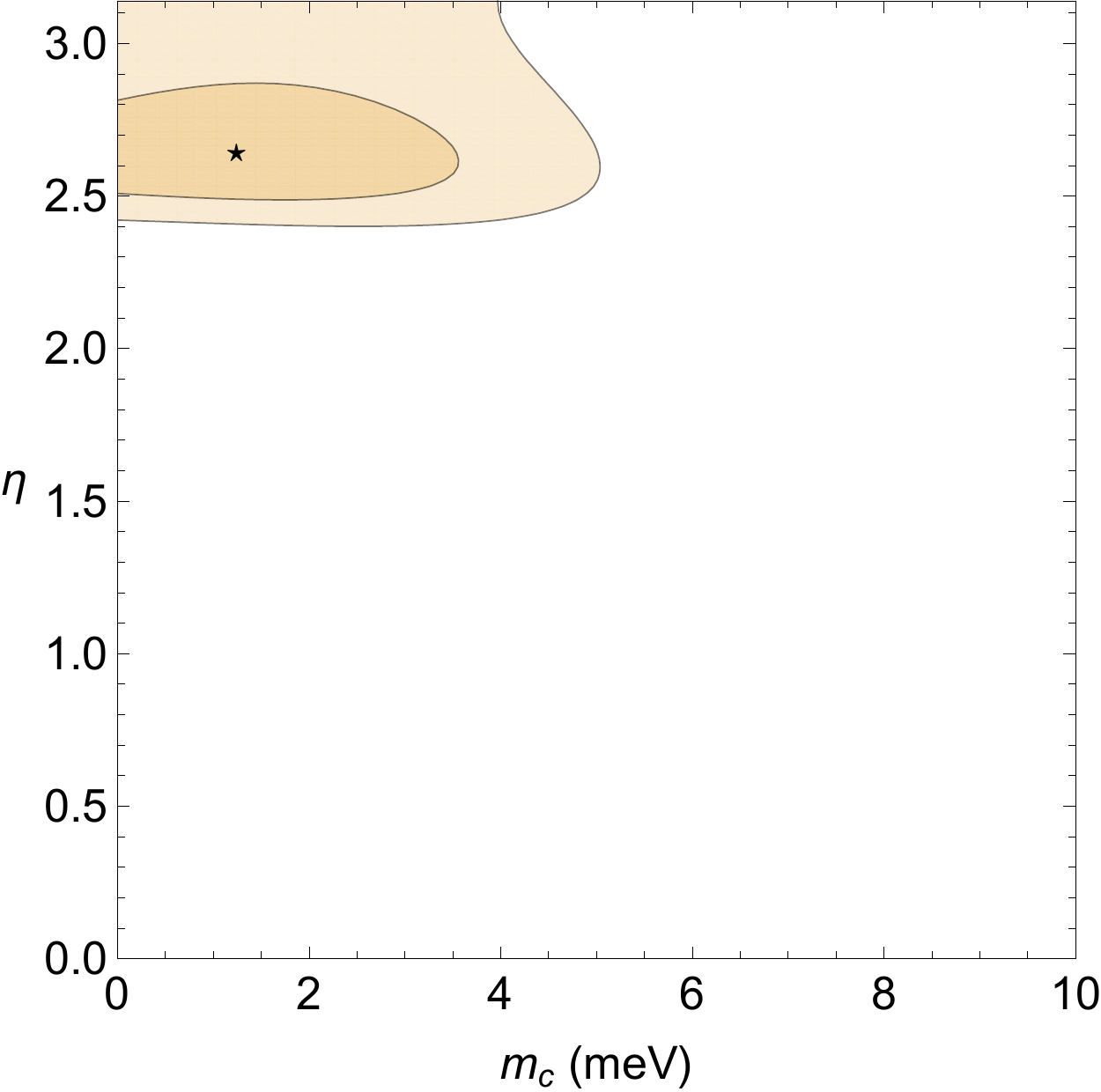}
		\caption{CSD(4), $ \xi = \eta $}
	\end{subfigure}\vspace{0.5cm}
	\begin{subfigure}{0.50\textwidth}
		\centering
		\includegraphics[scale=0.46]{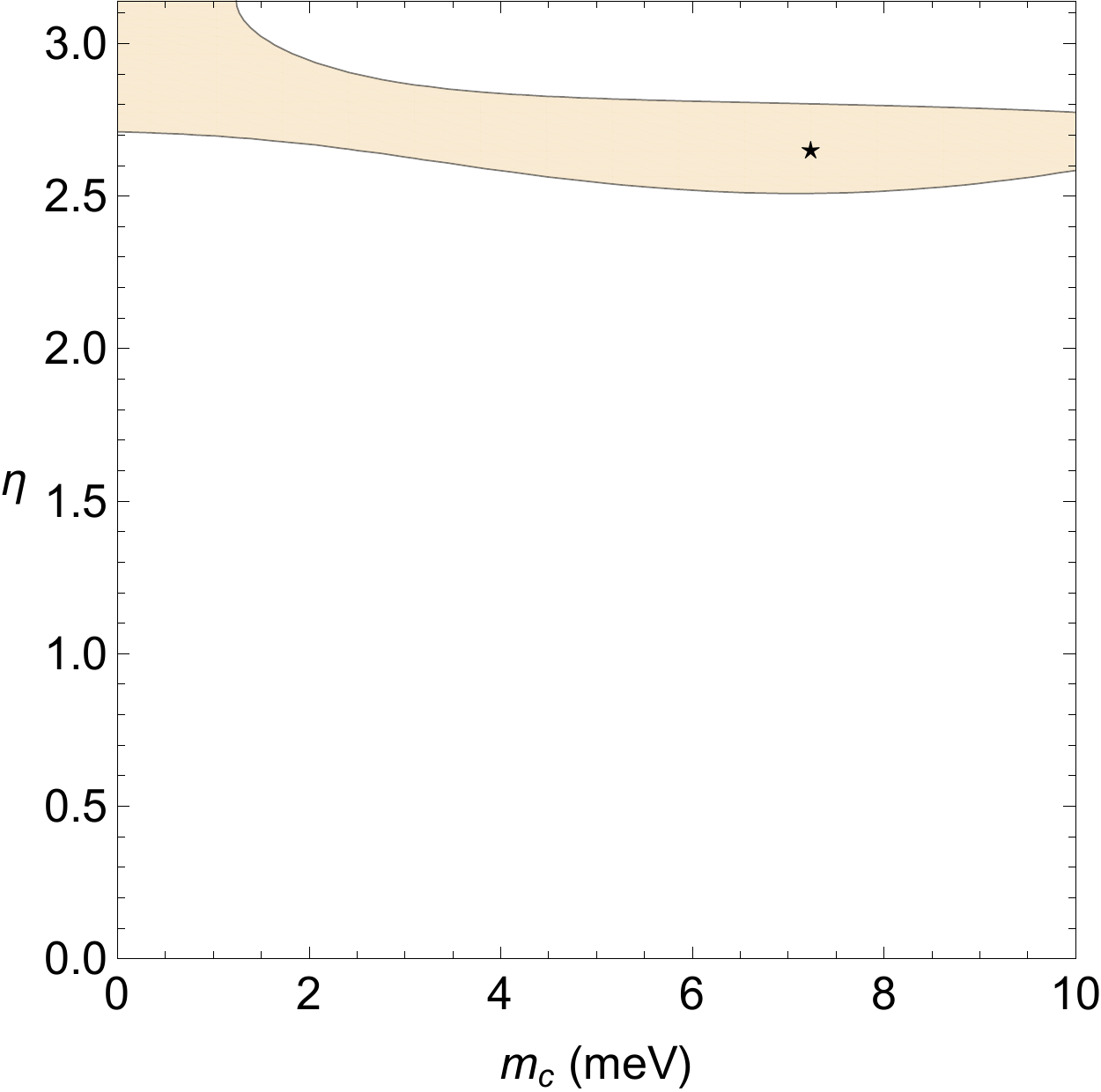}
		\caption{CSD(5), $ \xi = 0 $}
	\end{subfigure}%
	\hspace*{-2em}
	\begin{subfigure}{0.50\textwidth}
		\centering
		\includegraphics[scale=0.46]{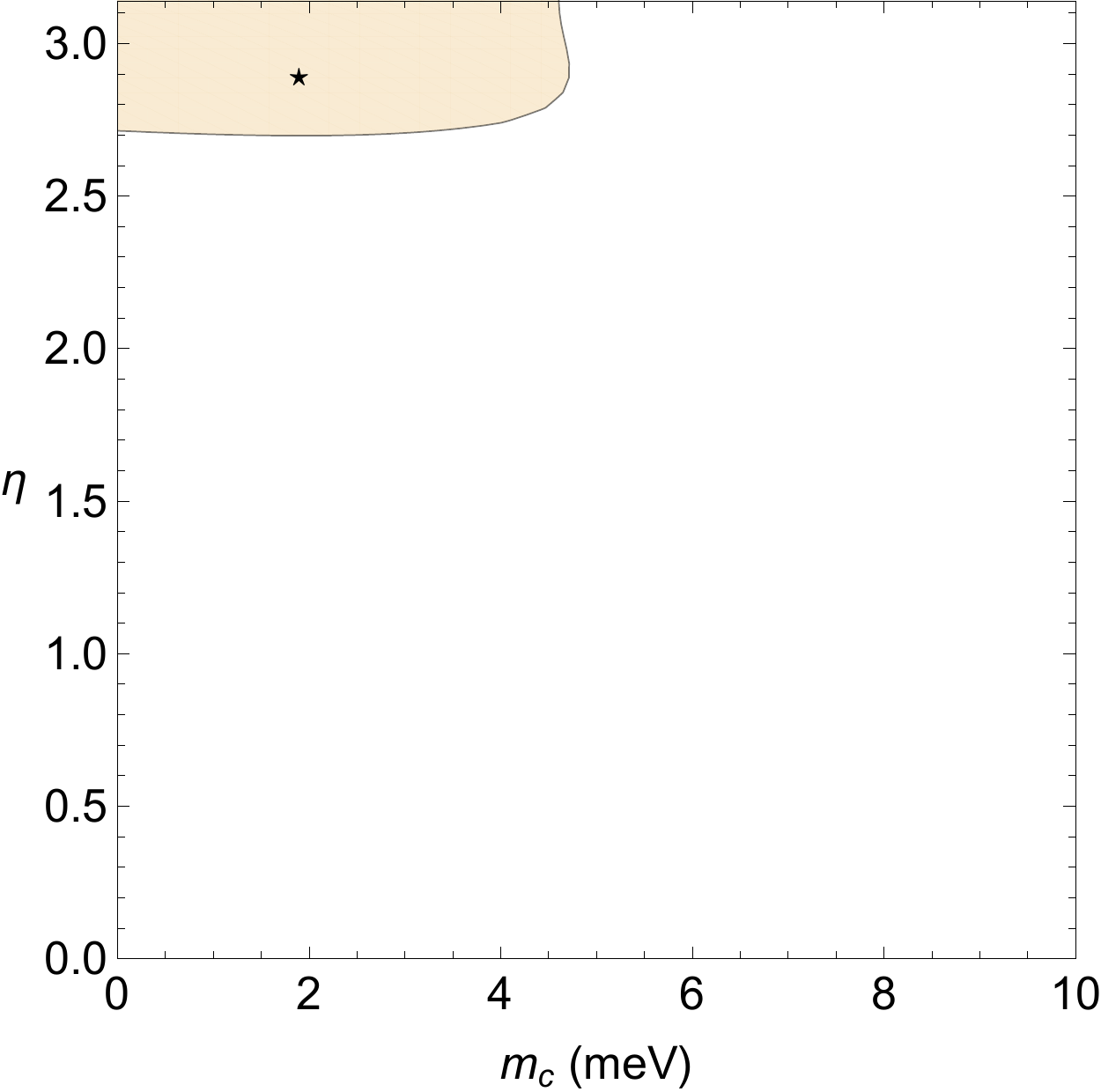}
		\caption{CSD(5), $ \xi = \eta $}
	\end{subfigure}
	\caption{Best fit $ \fit $ with respect to the phase $ \eta $ and third input neutrino mass $ m_c $. Note that the cases with $ \xi = 0 $ (left column) typically allow for a larger range of $ m_c $ values that give acceptable fits, when compared to the $ \xi = \eta $ cases. In these plots, the dark blue region corresponds to $\fit\leq 5$, while the surrounding regions correspond to $\fit\leq 20$ and $\fit\leq 50$. The best fit points are indicated by stars.}
	\label{fig:chisq_mceta_csd3-4}
\end{figure}

\begin{figure}[ht]
	\centering
	\includegraphics[scale=0.7]{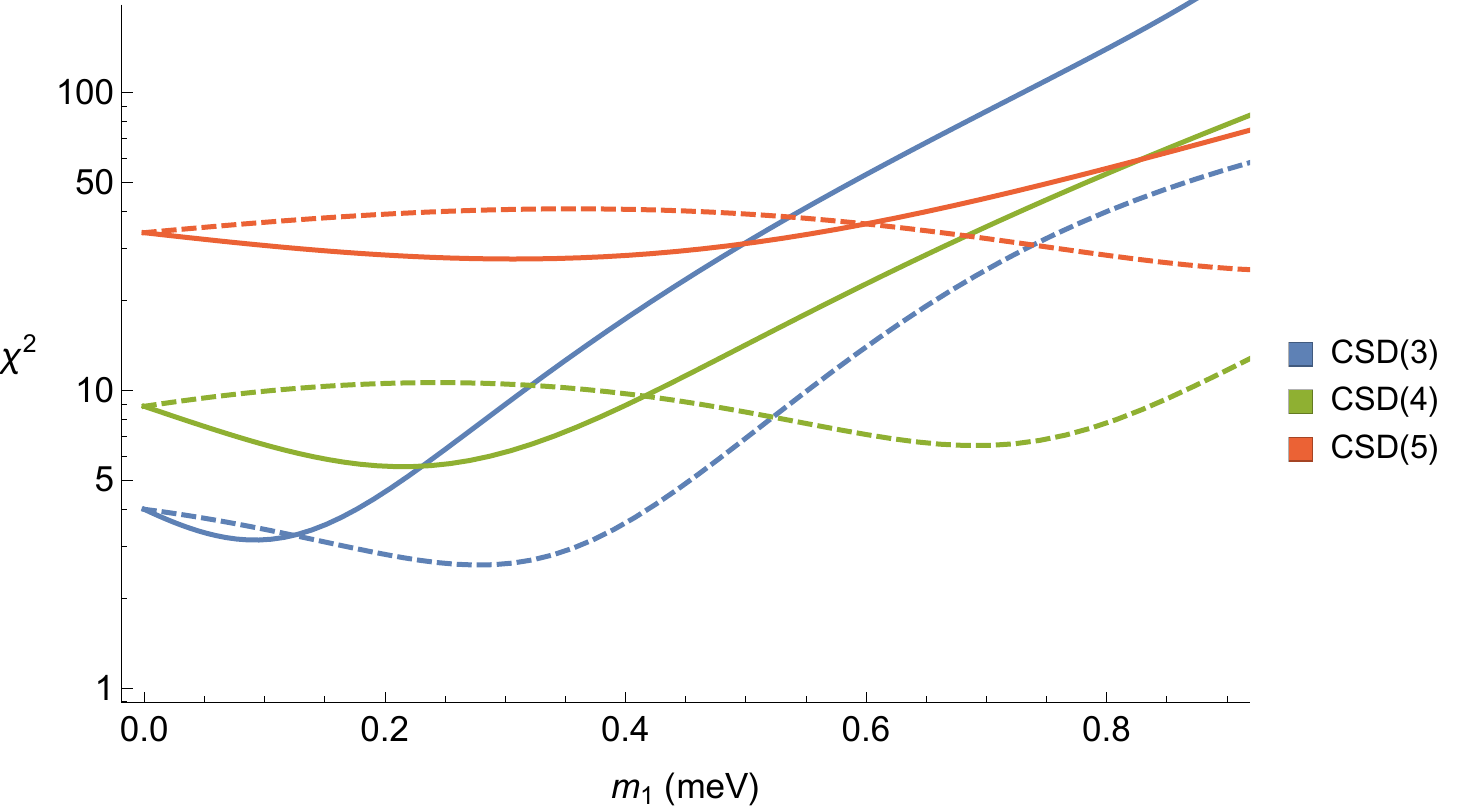}
	\caption{Best fit $\fit$ plotted with respect to the lightest neutrino mass $m_1$. 
	The dashed lines refer to the $\xi=0$ case, while the solid lines refer to the $\xi=\eta$ case.}
	\label{fig:chisqlog_m1_csd3-5}
\end{figure}

The variation of $ \fit $ with respect to the phase $ \eta $ and the third input neutrino mass $ m_c $ is shown in \figref{fig:chisq_mceta_csd3-4}. As in the case for models with two neutrinos, $ \eta $ is quite tightly constrained. Meanwhile, $ m_c $ typically has a larger range of acceptable values (particularly when $ \xi = 0 $), and does not appear strongly correlated with $ \eta $. Similarly the best fit values of the physical lightest neutrino mass $m_1$ lie in rather shallow minima of $\fit$ as shown in \figref{fig:chisqlog_m1_csd3-5}. 

In \figref{fig:inputs_m1_csd3-5} we show the variation of input parameters with lightest neutrino mass $m_1$, for the three best fit cases $ 3 \leq n \leq 5 $, while \figref{fig:mixingangles_m1_csd3-5} and \ref{fig:neutrinom_m1_csd3-5} show the dependence on $m_1$ for the predicted mixing angles and neutrino masses, respectively. In all these plots the dashed line is the $\xi=0$ case, while the solid line is the $\xi=\eta$ case.

\begin{figure}[ht]
	\centering
	\includegraphics[scale=0.75]{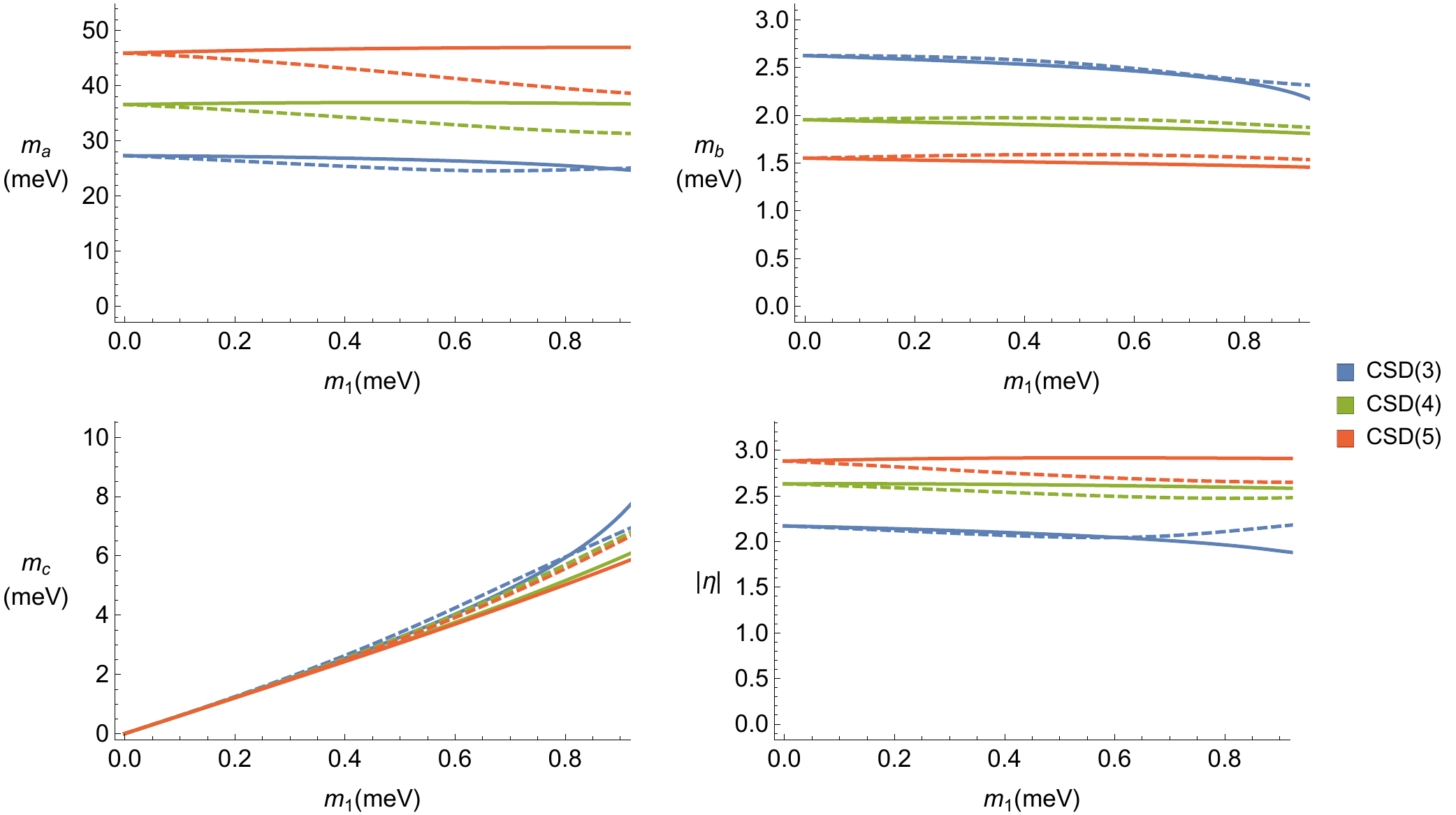}
	\caption{The variation with $m_1$ of the best fit input parameters. Note that $m_c$ and $m_1$ are closely correlated, while best fit $ m_{a,b} $ are not strongly affected by the introduction of a third neutrino. The dashed lines refer to the $\xi=0$ case, while the solid lines refer to the $\xi=\eta$ case.}
	\label{fig:inputs_m1_csd3-5}
\end{figure}

\begin{figure}[ht]
	\centering
	\includegraphics[scale=0.75]{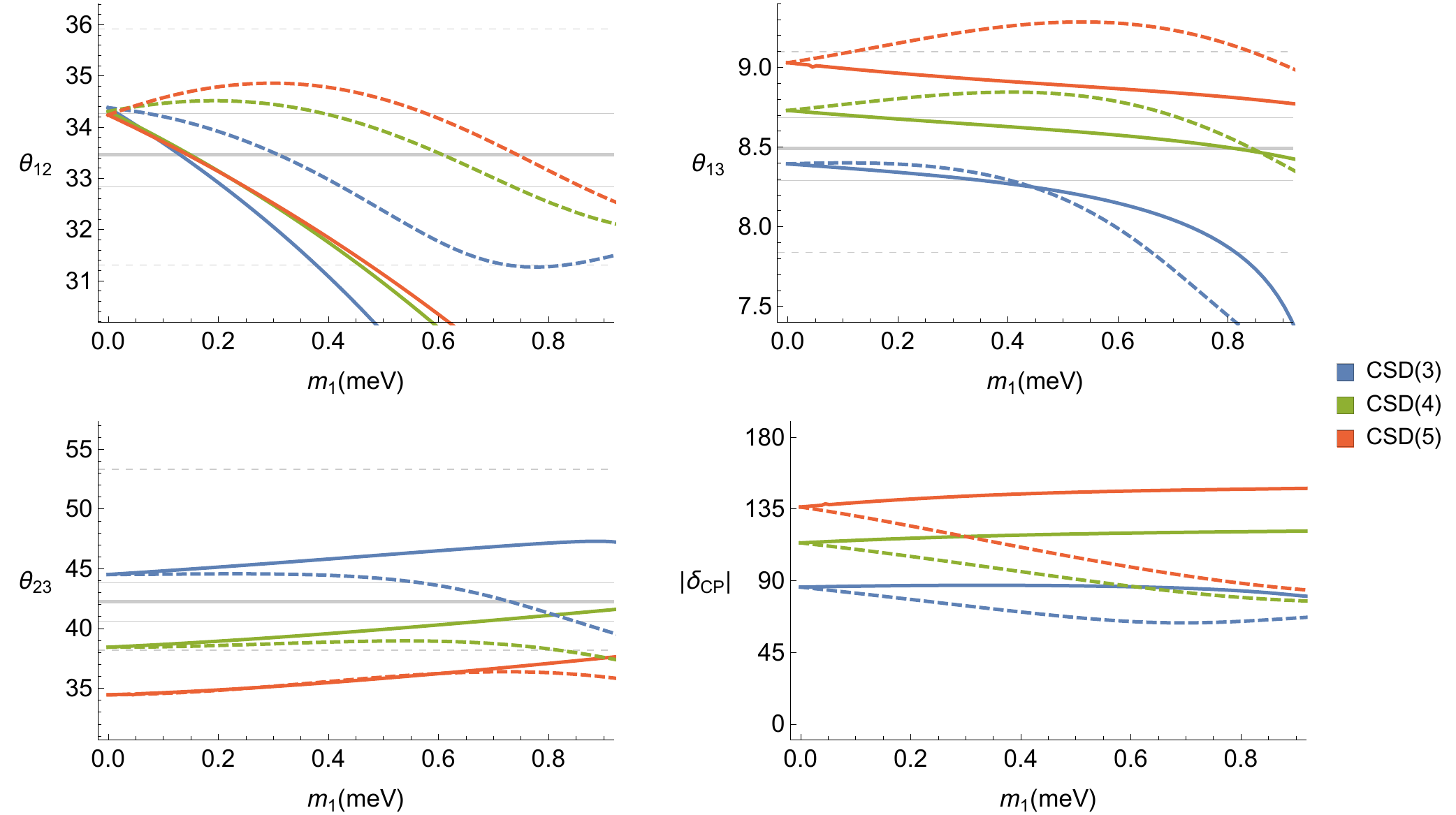}
	\caption{Predicted best fit mixing angles and the CP-violating phase plotted with respect to $m_1$. Note that the variation is mainly in $\theta_{12}$ when $ m_1 $ is small. The dashed lines refer to the $\xi=0$ case, while the solid lines refer to the $\xi=\eta$ case.}
	\label{fig:mixingangles_m1_csd3-5}
\end{figure}

\begin{figure}[ht]
	\centering
	\includegraphics[scale=0.75]{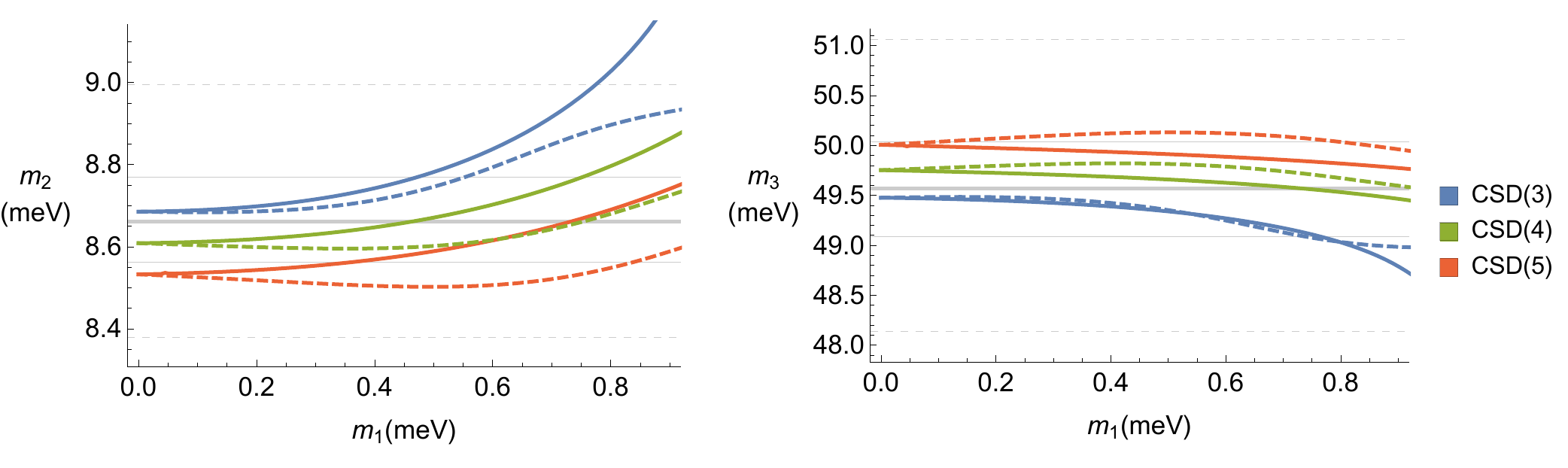}
	\caption{Second and third best fit neutrino masses plotted with respect to the lightest neutrino mass $m_1$. The horizontal gridlines drawn assume $ m_1 $ is negligible, such that $ m_2 \simeq \sqrt{\Delta m_{21}^2} $ and $ m_3 \simeq \sqrt{\Delta m_{31}^2} $.
	The dashed lines refer to the $\xi=0$ case, while the solid lines refer to the $\xi=\eta$ case.}
	\label{fig:neutrinom_m1_csd3-5}
\end{figure}

We observe that the choice of sub-subdominant phase $ \xi $ has a small effect on the value of the global minimum, but can noticeably shift its location in parameter space. Naturally the largest effect is on the best fit value and range of validity of $ m_c $, but it also contributes to interference between the three mass matrices in Eq.~\ref{eq:mnu}. The practical effect is that each of the three vacuum alignments contribute in varying amounts to each of the three PMNS mixing angles depending on the relative phase between the matrices, which can be seen particularly in \figref{fig:mixingangles_m1_csd3-5}, where the choice of $ \xi $ alters the shape of the variation of the mixing angles. 

More specifically, the addition of a third neutrino appears to most dramatically affect the solar angle $ \theta_{12} $, in contrast to the two neutrino model, where it is essentially constant. The physical neutrino masses in \figref{fig:neutrinom_m1_csd3-5} are comparatively far less sensitive to changes in $ \xi $.

\FloatBarrier
\subsection{CSD(3) with \secheadmath{\eta = 2\pi/3} and CSD(4) with \secheadmath{\eta = 4\pi/5}}
\label{sec:special}
It is interesting that the optimal fit for the CSD(3) with $ \fit = 2.59(3.14) $ corresponds to a choice of input phase $ |\eta| = 2.10(2.16) = 0.669\pi(0.682\pi) $, for the $ \xi=0 $ ($ \xi = \eta $) cases, respectively. Its closeness to the value $ 2\pi/3 $, independently of $ \xi $, is a compelling quality in favour of flavour models that predict additional $ \mathbb{Z}_{3N} $ symmetries, which tend to predict quantised phases as multiples of $ \pi/3 $. This motivates a $ \fit $ analysis with a fixed value of $ \eta = 2\pi/3 $, for a reduced input vector $ x = (m_a,m_b,m_c) $. The resulting input and output parameters for fixed $ \eta = 2\pi/3 $ are given in Table \ref{tab:csd3}. The best fits give $ \fit = 2.59(5.25) $, for the $ \xi=0 $ ($ \xi = \eta $) cases, respectively, marginally worse than in the case of unconstrained $ \eta $ fits which gave $ \fit = 2.59(3.14) $. 

In a model where $ \eta $ is fixed by symmetry, the excess degrees of freedom $ \nu =N-N_I$ are increased by one since $N_I$ is decreased by one. In CSD with two right-handed neutrinos, this brings the total to $ \nu = 3 $. Meanwhile, CSD with three right-handed neutrinos nominally has $ \nu = 0$. However, since the sub-subdominant phase $ \xi $ contributes only marginally to the goodness-of-fit, as has been seen in the analysis above, it can be fixed to some convenient value without affecting the conclusions drawn from the fit. Fixing both $ \xi $ and $ \eta $ gives $ \nu = 2 $ in the three-neutrino case. For both classes of models, we can attain $ \fit/\nu \sim 1.3-1.6 $ with CSD(3), which is slightly better than might be expected, although it should not be too suprising that one value of $n$ in CSD($n$) works better than the others (although it is encouraging that $n=3$ is a small number). The interpretation is also subject to the previous caveats about not including $\deltacp$ in the fit (which would increase $\nu$) and also our conservative treatment of non-Gaussian errors.

Turning to the other promising candidate, CSD(4), we see that, for $ \xi=0 $ ($ \xi = \eta $), we have $ \fit = 6.51(5.53) $ for $ |\eta| = 2.48(2.63)= 0.79\pi(0.84\pi) $, which is close to  $ 4\pi/5 $. Based on the work in \cite{King:2013hoa}, it is meaningful to examine the parameter space for a fixed phase $ \eta = \pm \frac{4\pi}{5}$ and $ \xi=0 $ or $\xi= \eta $.%
\footnote{Note that the first paper in \cite{King:2013hoa} involved  $ \xi=0 $ while the second paper used $ \xi = \eta $. However in such realistic models the charged lepton corrections also play a role.}
$ \fit $-minimisation yields $ \fit = 7.20(14.7) $ with corresponding input and output parameters given in Table \ref{tab:csd4}.

\begin{table}[ht]
\renewcommand{\arraystretch}{1.2}
\centering
\footnotesize
\begin{tabular}{ c | c | c c |}
\cline{3-4}
\multicolumn{2}{c |}{}								& $\xi = 0$ 	& $ \xi = \eta $ \\ \hline
\multicolumn{1}{| c |}{\multirow{3}{*}{Input}} 	& $m_a$ (meV) 		& 25.9		& 26.7\\
\multicolumn{1}{| c |}{}				& $m_b$ (meV) 		& 2.60		& 2.64\\
\multicolumn{1}{| c |}{}				& $m_c$ (meV)	 		& 1.80		& 0.88\\ \hline \cline{1-4}
\multicolumn{1}{| c |}{\multirow{7}{*}{Output}}	& $m_1$ (meV) 		& 0.29		& 0.14\\
\multicolumn{1}{| c |}{}				& $m_2$ (meV) 		& 8.71		& 8.63\\
\multicolumn{1}{| c |}{}				& $m_3$ (meV) 		& 48.2		& 49.7\\
\multicolumn{1}{| c |}{}				& $\theta_{12}$ ($^\circ$) 	& 32.1		& 33.3\\
\multicolumn{1}{| c |}{}				& $\theta_{13}$ ($^\circ$) 	& 8.74		& 8.54\\
\multicolumn{1}{| c |}{}				& $\theta_{23}$ ($^\circ$) 	& 46.2		& 45.8\\
\multicolumn{1}{| c |}{}				& $\deltacp$ ($^\circ$) 	& 90.2		& 89.1\\
\hline
\end{tabular}
\caption{Best-fit input and output values corresponding to $\fit = 2.59(5.25) $ for $\xi=0 $ ($\xi = \eta $) for a CSD(3) with a fixed input phase $\eta = 2\pi /3 $. }
\label{tab:csd3}	
\end{table}

\begin{table}[ht]
\renewcommand{\arraystretch}{1.2}
\centering
\footnotesize
\begin{tabular}{| c | c | c c |}
\cline{3-4}
\multicolumn{2}{c|}{}								& $\xi = 0$ 	& $ \xi = \eta $ \\ \hline
\multicolumn{1}{| c |}{\multirow{3}{*}{Input}} 	& $m_a$ (meV) 		& 33.0		& 35.4\\
\multicolumn{1}{| c |}{}				& $m_b$ (meV) 		& 1.94		& 1.99\\
\multicolumn{1}{| c |}{}				& $m_c$ (meV) 		& 4.42		& 1.60\\ \hline \cline{1-4}
\multicolumn{1}{| c |}{\multirow{7}{*}{Output}}	& $m_1$ (meV) 		& 0.66		& 0.26\\
\multicolumn{1}{| c |}{}				& $m_2$ (meV) 		& 8.65		& 8.49\\
\multicolumn{1}{| c |}{}				& $m_3$ (meV) 		& 49.7		& 50.2\\
\multicolumn{1}{| c |}{}				& $\theta_{12}$ ($^\circ$) 	& 33.5		& 32.7\\
\multicolumn{1}{| c |}{}				& $\theta_{13}$ ($^\circ$) 	& 8.68		& 9.05\\
\multicolumn{1}{| c |}{}				& $\theta_{23}$ ($^\circ$) 	& 38.2		& 41.3\\
\multicolumn{1}{| c |}{}				& $\deltacp$ ($^\circ$) 	& 93.6		& 112\\
\hline
\end{tabular}
\caption{Best-fit input and output values corresponding to $ \fit = 7.20(14.7)$ for $\xi=0 $ ($ \xi = \eta $) for a CSD(4) with a fixed input phase $\eta = 4\pi /5 $. }
\label{tab:csd4}
\end{table}

\FloatBarrier
\section{The link between \secheadmath{\deltacp} and leptogenesis in CSD(\secheadmath{n})}
\label{link}
Leptogenesis \cite{DiBari:2012fz} is a leading candidate for the origin of matter-antimatter asymmetry in the universe. In the original form of CSD, the columns of the Dirac mass matrix in the flavour basis were orthogonal to each other and consequently the CP asymmetries for cosmological leptogenesis vanished \cite{Antusch:2006cw,King:2006hn}. Following the subsequent observation that leptogenesis also vanished for a range of other family symmetry models \cite{Jenkins:2008rb}, this undesirable feature was eventually understood \cite{Choubey:2010vs} to be a general consequence of see-saw models with form dominance \cite{Chen:2009um} (i.e. in which the columns of the Dirac mass matrix in the flavour basis are proportional to the columns of the PMNS mixing matrix).

In the case for CSD($n$), leptogenesis does not vanish since the columns of the Dirac mass matrix in the flavour basis are not orthogonal. To be precise, $(m^D_{\rm atm})^T=(0,a,a)$ and $(m^D_{\rm sol})^T=(b,nb,(n-2)b)$ from Eq.~\ref{sol} are not orthogonal for $n>1$.%
\footnote{Note that the orginal CSD($n=1$) case satisfies form dominance since $(0,a,a).(b,b,-b)=0$. Hence leptogenesis vanishes in this case. However CSD(1) is excluded due to observed reactor angle and we find $\fit\sim 500$.}
Interestingly, since the see-saw mechanism in CSD($n$) with two right-handed neutrinos only involves a single phase $\eta = \arg(b^2/a^2)$, both the leptogenesis asymmetries and the neutrino oscillation phase $\deltacp$ must necessarily originate from $\eta$, providing a direct link between the two CP violating phenomena in this class of models, as follows.

The produced baryon asymmetry $Y_B$ from leptogenesis in two right-handed neutrino models with CSD($n$) satisfies, following the arguments in \cite{Antusch:2006cw},
\begin{equation}\label{eq:BAU}
Y_B \propto \pm\sin \eta \: ,
\end{equation}
where the ``$+$'' sign applies to the case $M_{\rm atm}  \ll M_{\rm sol}$ and the ``$-$'' sign holds for the case $M_{\rm sol} \ll M_{\rm atm}$. Since the observed baryon asymmetry $Y_B$ is positive, it follows that, for $M_{\rm atm}  \ll M_{\rm sol}$, we must have $\sin \eta$ to be positive, while for $M_{\rm sol} \ll M_{\rm atm}$ we must have $\sin \eta$ to be negative. We have seen that for CSD($n$) positive $\eta$ is associated with negative $\deltacp$ and {\it vice versa}. Although the global fits do not distinguish the sign of $\eta$, the present hint that $\deltacp \sim -\pi/2$ would require positive $\eta$, then in order to achieve positive $Y_B$ we require $M_{\rm atm}  \ll M_{\rm sol}$, corresponding to ``light sequential dominance'', depicted in Fig. \ref{SM}, as considered in the two right-handed neutrino analysis in \cite{Antusch:2011nz}.

The above link between CP violation in flavour dependent leptogenesis and neutrino oscillation for models with sequential dominance was observed in \cite{Antusch:2006cw}, although with only one leptogenesis phase the conclusions are identical to those obtained in the flavour independent or ``vanilla'' case \cite{King:2002qh}. Our discussion here generalises that of CSD(2) which involves two texture zeroes\cite{Antusch:2011ic}. Here we find a link for CSD($n$), even without two texture zeroes, due to the appearance of only a single phase $\eta$ in the see-saw mechanism, for the case of two right-handed neutrinos, where $\eta$ is identified as both the leptogenesis phase in Eq.~\ref{eq:BAU} and the phase appearing in the neutrino mass matrix in Eq.~\ref{eq:mnu2}.

The above conclusions remain approximately true when a third almost decoupled right-handed neutrino is introduced.
As discussed in \cite{Antusch:2006cw}, the relative size of the additional contribution to the CP asymmetry when a third neutrino is present is $ \mathcal{O}(m_c/m_b) \sim 0.1 $.
A third right-handed neutrino is necessary in the realistic Pati-Salam models based on CSD(4) in \cite{King:2013hoa}. In these models the new phase is either given by $\xi=0$ or $\xi=\eta$, so no new leptogenesis phase appears. However the mechanism for leptogenesis is necessarily quite different in these models, since the lightest right-handed neutrino of mass $M_{\rm atm}$ is too light to generate successful leptogenesis in its decays. Instead one must rely on the decays of the second lightest right-handed neutrino of mass $M_{\rm sol}$ as in $SO(10)$-inspired leptogenesis \cite{DiBari:2014eya}. It would be interesting to discuss this in more detail in a future publication \cite{DiBari:2015oca}.

\FloatBarrier
\section{Conclusion}
\label{conclusion}
We have performed an analysis on a class of CSD($n$) models in which, in the flavour basis, two right-handed neutrinos are dominantly responsible for the ``atmospheric'' and ``solar'' neutrino masses with Yukawa couplings to $(\nu_e, \nu_{\mu}, \nu_{\tau})$ proportional to $(0,1,1)$ and $(1,n,n-2)$, respectively, where $n$ is a positive integer. The $ \fit $ measure offers a flexible and robust way to examine parameter space for a given form of neutrino mass matrix, and the relative strength of fit to experimental data. However the treatment and interpretation of this statistical test is subject to a number of subtleties and caveats which we have discussed already. These include: the non-Gaussianity of the ``data'' (taken from a particular global analysis); the decision about whether to include $\deltacp$ as a measured ``data'' point; the treatment of parameters which do not significantly affect the fit, namely $\xi$, and those which are constrained to be small on theoretical grounds, namely $m_c$; and which parameters are regarded as being fixed by theory such as $n$ of CSD($n$) and the phase $\eta$ which may be fixed by a discrete symetry. We have tried to be very conservative at each step, and have clearly discussed the interpretation taking into account our conservative approach. For example, we have taken the smallest asymmetric errors when dealing with non-Gaussian distributions. We have also carefully discussed the relevant $\nu = N-N_I$ with which the $\fit$ should be compared, in each case, and seen that the interpretation is non-trivial.

With the above caveats in mind, we can say that for just two right-handed neutrinos, we cautiously find good agreement with experiment for CSD(3) and CSD(4), leading to accurate predictions for mixing angles, with these two cases being distinguished by their differing predictions for the atmospheric angle of $\theta_{23}\approx 45^{\circ}$ and $\theta_{23}\approx 39^{\circ}$, respectively. We find it encouraging that the entire PMNS matrix can be so accurately fitted in terms of just one input parameter, namely the relative phase $\eta$, with the input mass parameters $m_a$ and $m_b$ essentially determining the neutrino mass-squared differences $ \Delta m^2_{21} $, $ \Delta m^2_{31} $. In some models the phase $\eta$ itself may be fixed to some small set of discrete values consistent with the preferred values extracted from the fit.

We have also carefully studied the perturbing effect of a third ``decoupled'' right-handed neutrino, leading to a bound on the lightest physical neutrino mass $m_1\simlt 1$ meV for the viable cases, corresponding to the robust prediction of a normal neutrino mass hierarchy. The best fit model in this case is shown to be CSD(3), for which we find $ \fit = 2.59(3.14) $ with fixed sub-subdominant phase $ \xi=0 \ (\xi=\eta) $. CSD(3) has the added desirable property that $\eta$ is preferred to be close to $2\pi/3$, which can be naturally predicted from flavour models with a $ \mathbb{Z}_{3N} $ symmetry. Another leading contender is CSD(4) for which we find $ \fit = 6.51(5.53) $ with fixed sub-subdominant phase $ \xi=0 \ (\xi=\eta) $. For the case $ \xi=0 $, CSD(4) prefers $\eta$ to be close to $4\pi/5$, which can be naturally predicted from flavour models with a $ \mathbb{Z}_{5N} $ symmetry. However such realistic flavour models will in general also include (small) charged lepton corrections which will modify the fits presented here. CSD(5) is less favoured with $ \fit \approx 25\!-\!30$. The present analysis confirms that CSD(2), as well as CSD($ n>5 $) are disfavoured, the latter being a new result.

The analysis is further complicated for larger $n$ by the increasing influence of the sub-subdominant matrix proportional to $ m_c $ (which correlates very closely with the lightest neutrino mass). However here we restrict ourselves to the CSD framework where we demand that the third right-handed neutrino be approximately decoupled from the see-saw mechanism, since otherwise all predictivity is lost. Ultimately, while it is not perhaps suprising a reasonable fit can be found for some $ n $, unexpectedly only a handful of alignments ($ n = 3,4 $ and possibly $ n=5 $) give a good agreement with data, all with reasonably small $ n $.

Furthermore, the predictions for mixing angles and masses are quite sharp, particularly for the viable choices of CSD(3) and CSD(4), and could feasibly be ruled out by more precise measurements of mixing angles and CP phase in future neutrino experiments.

Finally we have seen that, with just two right-handed neutrinos in CSD($n$), there is a direct link between the oscillation phase $\deltacp $ and leptogenesis. This is because there is only one phase $\eta$ appearing in the see-saw mass matrices. Hence $\eta$ is identified as both the leptogenesis phase in Eq.~\ref{eq:BAU} and the phase in the neutrino mass matrix in Eq.~\ref{eq:mnu2}. For a given ordering of right-handed neutrino masses, the sign of $\eta$ is fixed by the requirement that the observed baryon asymmetry is positive. For instance, if $M_{\rm atm}  \ll M_{\rm sol}$, then positive baryon asymmetry requires that $\eta$ is also positive. A positive $\eta$ in the neutrino mass matrix implies that $\deltacp $ is negative, in agreement with the current hint that $\deltacp \sim -\pi/2$. It is interesting that, for positive $\eta$, the global analyses for CSD($3$) or CSD($4$) with two right-handed neutrinos predict values of $\deltacp \approx -92.2^{\circ}$ or $\deltacp \approx -120^{\circ}$, respectively, with similar predictions when a third approximately decoupled right-handed neutrino is included.

In summary, the $ \fit $ fit of the input parameters to measured parameters presented here confirms and quantifies the earlier claims that see-saw models based on CSD($3$) and CSD($4$) successfully describe neutrino masses and lepton mixings. Although the interpretation of the $\fit$ values is subject to discussion, we have tried to be clear and open as well as conservative when presenting the results. In the case of two right-handed neutrinos, a single phase $\eta$ controls leptogenesis and predicts all the mixing angles and phases in the PMNS matrix. Overall, it is encouraging that good fits for these two cases can be achieved, subject to the above discussed caveats, for simple values of the phase $\eta = 2\pi/3$ or $\eta = 4\pi/5$, where such quantised values of the phases could arise due to Abelian $ \mathbb{Z}_{3N} $ or $ \mathbb{Z}_{5N} $ discrete symmetries, together with a non-Abelian $A_4$ family symmetry. In such models the CSD($n$) structures emerge from $A_4$ breaking flavons with vacuum alignments proportional to $(1,n,n-2)^T$, arising from othogonality conditions. However, in such realistic models, we should in general expect the results presented here to be modified somewhat due to further small effects arising from charged lepton mixing and renormalisation group corrections. In conclusion, the see-saw mechanism with CSD($n$) represents a predictive and rather successful approach to understanding lepton mixing with a direct link to leptogenesis.

\section*{Acknowledgements}
We would like to thank Alex Merle for useful correspondence.
We acknowledge support from the European Union FP7 ITN-INVISIBLES (Marie Curie Actions, PITN- GA-2011-289442). 

\appendix

\section{\secheadmath{\fit} near the global minimum}
\label{sec:chisqinput}

\figref{fig:chisq_mamb_csd3-4} shows the best fit $ \fit $ with respect to the two input masses $ m_a $ and $ m_b $ for a two neutrino model for the most physically interesting cases of CSD(3) and CSD(4). It is clear from the contours that both input masses are quite tightly constrained. Any fit that gives $ \fit < 50 $ will correspond to a deviation from the best fit value of no more than 10-15\%. It is also confirmed that the addition of a third right-handed neutrino does not significantly alter the best-fit or the spread of $ m_a $ and $ m_b $, as $ m_c $ is small (as is required by CSD). This lends validity to our assertion that the two physical neutrino masses $ m_{2,3} $ are largely derived from the input masses $ m_{a,b} $, leaving (in the two neutrino case) only a single phase $ \eta $ which controls the detailed prediction of the PMNS matrix.

\begin{figure}[ht]
	\centering
	\begin{subfigure}{0.45\textwidth}
		\centering
		\includegraphics[scale=0.5]{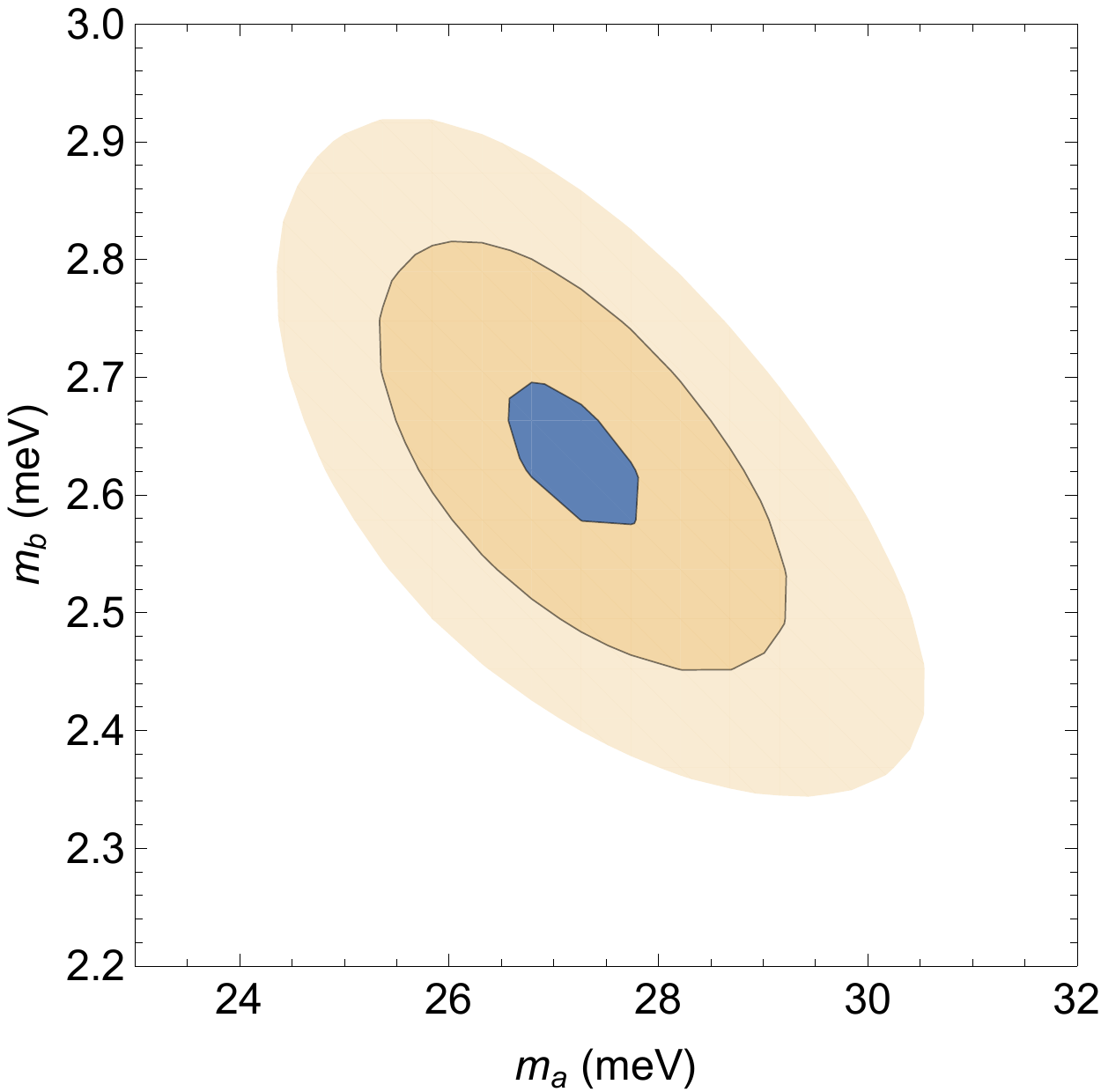}
		\caption{CSD(3)}
	\end{subfigure}%
	\begin{subfigure}{0.45\textwidth}
		\centering
		\includegraphics[scale=0.5]{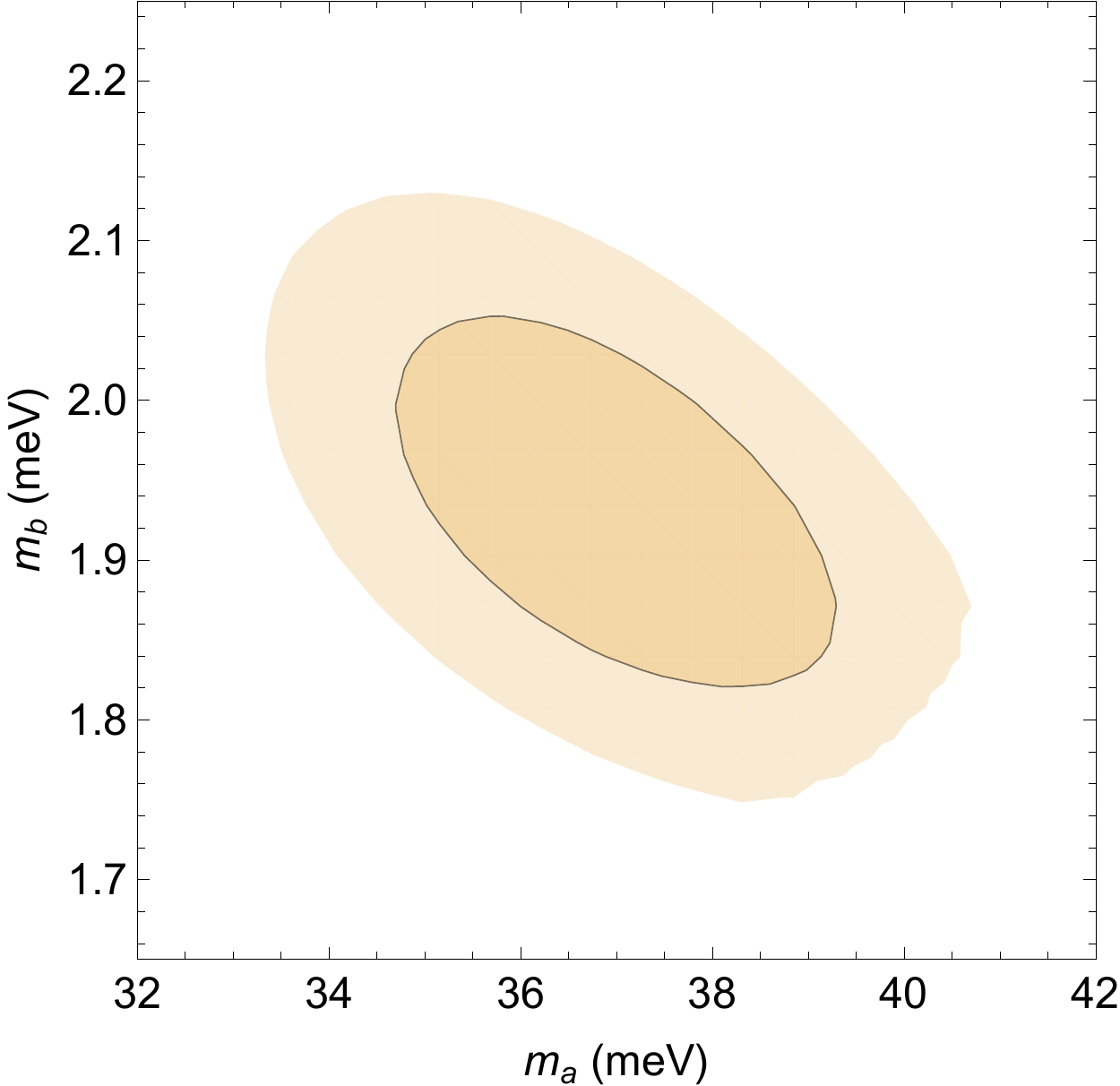}
		\caption{CSD(4)}
	\end{subfigure}
	\caption{Variation of $ \fit $ with input masses $ m_a $ and $ m_b $ for CSD(3) and CSD(4). The dark blue region corresponds to $\fit\leq 5$, while the surrounding regions correspond to $\fit\leq 20$ and $\fit\leq 50$.}
	\label{fig:chisq_mamb_csd3-4}
\end{figure}

The validity of $ \fit $ as a test-statistic depends not only on its ability to measure the numerical fit to data but also how reliable its behaviour is in the neighbourhood of a purported global minimum. We find that once we have constrained our parameter space to exclude degenerate minima (corresponding to $ \pm \eta $), $ \fit $ as defined in \secref{sec:chisq} is typically well-behaved near the observed minimum. \figref{fig:chisq_input_csd3-4} shows, for CSD(3) and CSD(4), the variation of $ \fit $ in this region. Specifically, it plots the lower envelope of $ \fit $ evaluated for $ 10^6 $ vectors in parameter space $ (m_a,m_b,m_c,\eta) $, chosen randomly. This means this envelope is not subject to any systematic errors due to any minimising algorithm. Other CSD($n$) alignments do not demonstrate any different behaviours to those observed in these graphs. 

The shape of the curves for $ m_a $, $ m_b $, and $ \eta $ show clearly defined minima, while the range of low-$\fit$ values is comparatively wider and includes $ m_c = 0 $. Nevertheless, although $ m_c $ may take a large range of values and produce reasonably good $ \fit $ fits, it appears to have a single minimum region -- the global and any local minima are the same. We safely fix the sub-subdominant phase $ \xi $ in \figref{fig:chisq_input_csd3-4}, which has a negligible effect on the position and nature of the minimum.

\begin{figure}[ht]
	\centering
	\begin{subfigure}{\textwidth}
		\centering
		\includegraphics[scale=0.7]{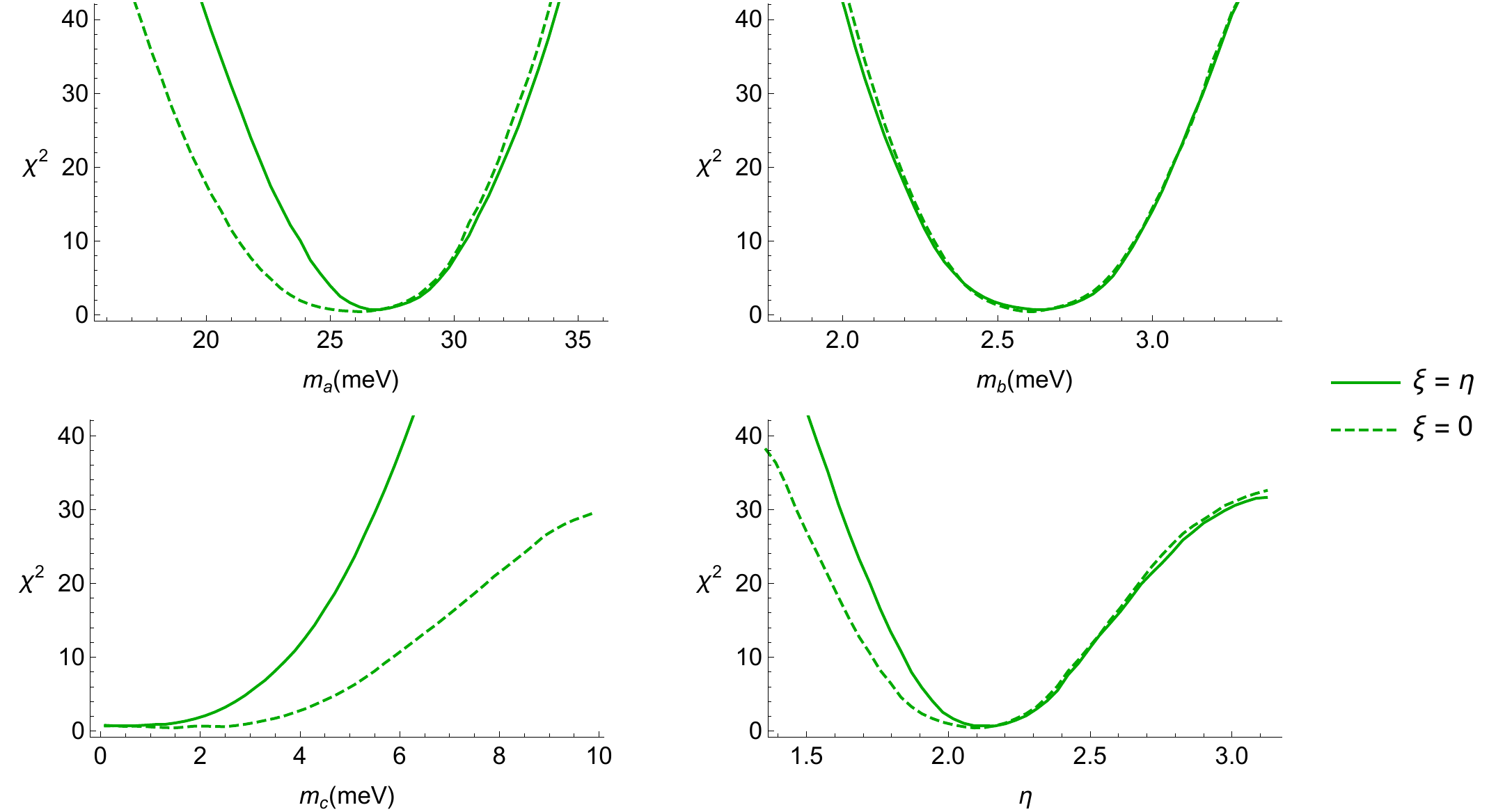}
		\caption{CSD(3)}
	\end{subfigure}\vspace{1cm}
	\begin{subfigure}{\textwidth}
		\centering
		\includegraphics[scale=0.7]{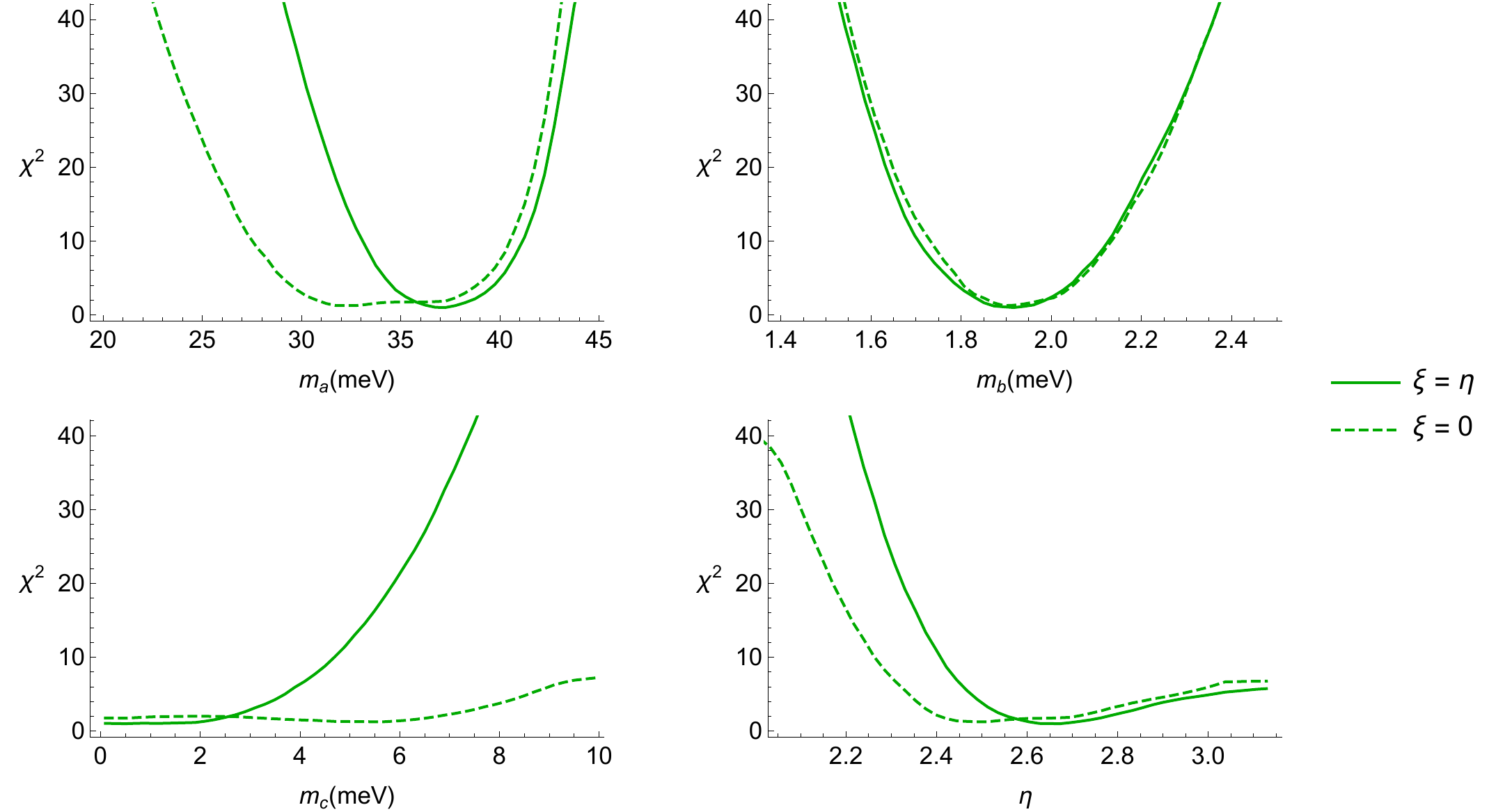}
		\caption{CSD(4)}
	\end{subfigure}
	\caption{Lower envelope of best fit $ \fit $ in the neighbourhood of the global best fit of input parameters $ m_a $, $ m_b $, $ m_c $, and $ \eta $ for CSD(3) and CSD(4), with fixed $ \xi $.}
	\label{fig:chisq_input_csd3-4}
\end{figure}

\FloatBarrier

\end{document}